\documentclass[astrosymb, times, trackchanges]{aastex631}

\accepted{\today}

\usepackage{relsize}
\newcommand{\lasp}{Laboratory for Atmospheric and Space Physics, University of Colorado, 600 UCB, Boulder, CO 80309, USA}
\newcommand{\casa}{Center for Astrophysics and Space Astronomy, University of Colorado, 593 UCB, Boulder, CO 80309}
\newcommand{\cuboulder}{Department of Astrophysical and Planetary Sciences, University of Colorado, Boulder, CO 80309, USA}

\newcommand{\zenodorepo}{ \url{https://doi.org/10.5281/zenodo.6909473}}
\newcommand{\lsfurl}{\url{https://www.stsci.edu/hst/instrumentation/stis/performance/spectral-resolution}}

\shorttitle{FUMES III: Ultraviolet and Optical Variability of M Dwarf Chromospheres}
\shortauthors{Duvvuri et al.}
\graphicspath{{./}{figures/}}

\begin{document}

\title{FUMES III: Ultraviolet and Optical Variability of M Dwarf Chromospheres}

\author[0000-0002-7119-2543]{Girish~M.~Duvvuri}
\affiliation{\cuboulder}
\affiliation{\casa}
\affiliation{\lasp}

\author[0000-0002-4489-0135]{J.~Sebastian~Pineda}
\affiliation{\cuboulder}
\affiliation{\lasp}

\author[0000-0002-3321-4924]{Zachory~K.~Berta-Thompson}
\affiliation{\cuboulder}
\affiliation{\casa}

\author[0000-0002-1002-3674]{Kevin~France}
\affiliation{\lasp}
\affiliation{\cuboulder}
\affiliation{\casa}

\author[0000-0002-1176-3391]{Allison~Youngblood}
\affiliation{Exoplanets and Stellar Astrophysics Lab, NASA Goddard Space Flight Center, Greenbelt, MD 20771, USA}


\begin{abstract}
We obtained ultraviolet and optical spectra for 9 M~dwarfs across a range of rotation periods to determine whether they showed stochastic intrinsic variability distinguishable from flares. The ultraviolet spectra were observed during the Far Ultraviolet M~Dwarf Evolution Survey \emph{Hubble~Space~Telescope} program using the Space Telescope Imaging Spectrograph. The optical observations were taken from the Apache Point Observatory 3.5-meter telescope using the Dual Imaging Spectrograph and from the Gemini South Observatory using the Gemini Multi-Object Spectrograph. We used the optical spectra to measure multiple chromospheric lines: the Balmer series from H$\alpha$ to H$10$ and the \ion{Ca}{2}~H and K lines. We find that after excising flares, these lines vary on the order of $1-20\%$ at minute-cadence over the course of an hour. The absolute amplitude of variability was greater for the faster rotating M~dwarfs in our sample. Among the 5 stars for which measured the weaker Balmer lines, we note a tentative trend that the fractional amplitude of the variability increases for higher order Balmer lines. We measured the integrated flux of multiple ultraviolet emission features formed in the transition region: the \ion{N}{5}, \ion{Si}{4} and \ion{C}{4} resonance line doublets, and the \ion{C}{2} and \ion{He}{2} multiplets. The signal-to-noise (S/N) ratio of the UV data was too low for us to detect non-flare variability at the same scale and time cadence as the optical. We consider multiple mechanisms for the observed stochastic variability and propose both observational and theoretical avenues of investigation to determine the physical causes of intrinsic variability in the chromospheres of M~dwarfs.
\end{abstract}



\section{Introduction} \label{sec:intro}

Stellar activity is a catch-all term for various behaviors in the stellar atmosphere produced by non-thermal heating processes,  processes which create distinct layers above the photospheres of most stars: the chromosphere, transition region, and corona \citep{Linsky:2017}. For low-mass stars, nearly all the high-energy emission (from wavelengths $< 2000 \, $\r{A}) is a product of stellar activity, varying over time as the stellar magnetic and atmospheric structures interact. One category of magnetic activity is the presence of surface inhomogeneities (spots, faculae, plages) which contribute emission distinct from the rest of the photosphere on varying timescales: cycling in and out of view over the rotation period of the star which can be anywhere between a few hours and a year \citep{Newton:2016}; waxing and waning over weeks as the surface features erupt, evolve, and dissipate \citep{Basri:2020}; and growing or fading over decades as the distribution, frequency, and strength of the surface features are changed by the cycling of the stellar magnetic field \citep{Wilson:1968}. Another category is flaring, which is a seconds-to-hours burst of emission produced by magnetic reconnection events depositing energy. Flare emission can manifest as broadband enhancements to the optical, ultraviolet, and X-ray continua, radio synchrotron/gyrosynchrotron emission, and/or enhancements to emission lines that grow in amplitude, broaden, and sometimes blue/redshift during the flare \citep{Kowalski:2013, MacGregor:2021}. Finally, all forms of stellar activity are subject to evolution over the megayear timescales of angular momentum evolution \citep{Skumanich:1972}.

As the study of M dwarfs has been intensified by the interest in their terrestrial exoplanets, the time-variability of magnetic activity has been studied both for its effects on exoplanet observations \citep{Llama:2015, Rackham:2019} and for its physical impact on the formation, evolution, and retention of exoplanetary atmospheres \citep{Shields:2016}. This motivates studying the high-energy emission of low mass stars across a wide range of rotation periods, which implicitly correspond to a wide range of activity levels and ages.

The Far Ultraviolet M-dwarf Evolution Survey (FUMES) collected far ultraviolet (FUV) spectra between $1100 \, \textrm{--} \, 1736 \,$\r{A} of a sample of intermediate activity stars (whose stellar properties are listed in Table \ref{tab:fumes_sample}) to complement existing datasets and enable new studies of the rotational evolution of magnetic activity for M-dwarfs. \citet{Pineda:2021a}, FUMES I, measured the flux of FUV emission lines formed in the transition region to study the rotation-activity relation in this wavelength/structural regime. \citet{Youngblood:2021}, FUMES II, reconstructed the intrinsic Lyman-$\alpha$ emission for the FUMES sample and demonstrated that the wings of the Lyman-$\alpha$ line can be used to infer the density of the chromosphere. This work, FUMES III, complements the UV data with optical spectroscopic measurements of the Balmer series and \ion{Ca}{2} H and K emission lines to study the stochastic variability of the chromosphere and transition region on short timescales comparable to exoplanet observations. When we use the term ``stochastic", we are referring to brief changes that cannot be clearly identified as flares based on their amplitude or light curve structure. This analysis builds on existing literature examining the stochastic emission line variability of low mass stars in the ultraviolet \citep{Loyd:2014} and the optical \citep{Lee:2010, Kruse:2010, Medina:2022}, but extends it by applying similar analytical techniques to both of these disparate wavelength (and implicitly structural) regimes for a common sample of stars on similar timescales.

Section \S\ref{sec:optical_observations} discusses the optical observations and their reduction (\S\ref{sec:optical_reduction}), and the measurement of equivalent widths and creation of spectral line light curves  to look for variability (\S\ref{sec:optical_analysis}). Section \S\ref{sec:uv_analysis} describes our analysis of the UV data to create similar spectral line light curves for transition region emission lines. In Section \S\ref{sec:variability} we quantify the significance of observed variability and compare the variability of different lines to each other and for each line as a function of Rossby number. Finally in Section \S\ref{sec:conclusion} we summarize our findings, discuss the physical and observational implications of our work, and conclude by recommending some observational and theoretical work that could help determine the origin of the stochastic variability described in this study.


\begin{deluxetable}{cCCCccc}
\tablehead{\colhead{Name} & \colhead{Mass} & \colhead{Radius} & \colhead{$T_\textrm{eff}$} & \colhead{$P_\textrm{rot}$} & \colhead{Rossby Number} & \colhead{H$\alpha$ EW$_{\textrm{quiescent}}$} \\ \colhead{--} & \colhead{[$M_{\odot}$]} & \colhead{[$R_{\odot}$]} & [K] & \colhead{\textrm{[days]}} & \colhead{\textrm{--}} & \colhead{\r{A}}}
\startdata
HIP 112312 & 0.247_{-0.015}^{+0.018} & 0.688_{0.016}^{+0.015} & 3173_{-22}^{+25} & 2.355 & 0.0249 & 0.157\\
LP 247-13 & 0.495 \pm 0.013 & 0.49 \pm 0.02 & 3511_{-76}^{+79} & 1.289 & 0.0278 & 0.252\\
CD-35 2722 & 0.572 \pm 0.002 & 0.561 \pm 0.003 & 3727_{-6}^{+8} & 1.717 & 0.0459 & 0.299\\
HIP 17695 & 0.435_{-0.014}^{+0.011} & 0.502_{-0.007}^{+0.008} & 3393_{-21}^{+20} & 3.87 & 0.0708 & 0.226\\
HIP 23309 & 0.785_{-0.01}^{+0.009} & 0.932 \pm 0.014 & 3886_{-27}^{+28} & 8.6 & 0.379  & 0.558\\
GJ 410 & 0.557 \pm 015 & 0.549 \pm 0.024 & 3786_{-83}^{+89} & 14 & 0.385 & 0.289\\
GJ 49 & 0.541 \pm 0.015 & 0.534 \pm 0.023 & 3713_{-81}^{+86} & 18.6 & 0.472 & 0.278\\
G 249-11 & 0.237 \pm 0.006 & 0.255 \pm 0.01 & 3277_{-68}^{+71} & 52.76 & 0.571 & 0.152\\
LP 55-41 & 0.41 \pm 0.01 & 0.416  \pm 0.017 & 3412_{-73}^{+75} & 53.44 & 0.905 & 0.216
\enddata
\label{tab:fumes_sample}
\caption{These are the stellar parameters for the subset of the FUMES sample considered in this work. All parameters are from Table 2 of \citet{Pineda:2021a} as determined in \citet{Pineda:2021b}. The column titled ``H$\alpha$ EW$_{\textrm{quiescent}}$ lists the photospheric contribution to the H$\alpha$ equivalent width of the star, calculated using Equation 2 in \citet{Newton:2017} and the listed stellar masses as inputs.}
\end{deluxetable}

\section{Optical Spectra} \label{sec:optical_observations}
The bulk of the optical emission from low mass stars originates from the photosphere, which can be thought of as the ``surface" of the star, but the Balmer series and \ion{Ca}{2} H and K lines are prominent optical emission features that are formed higher in the magnetically heated chromosphere. The red tinge of the H$\alpha$ Balmer line coloring the solar limb near the endpoints of an eclipse is how the chromosphere was first identified and named as a distinct stellar atmospheric structure \citep{Lockyer:1868}. The Balmer series lines form through a combination of photoionization, recombination, and collisional excitation while the \ion{Ca}{2} H and K lines are collisionally dominated \citep{CramGiampapa:1987}. Observing multiple optical chromospheric emission lines and comparing them to the photospheric continuum traces the physics of magnetic heating in the chromosphere relative to the local thermodynamic equilibrium of the photosphere.

While most M dwarfs display chromospheric \ion{Ca}{2} H and K emission, only a fraction of these also show the H$\alpha$ line in emission \citep{StaufferHartmann:1986}. The least active M dwarfs show shallow H$\alpha$ absorption, and an increase in chromospheric density first deepens the H$\alpha$ absorption, then fills in the absorption feature with emission, and finally exceeds the photospheric continuum level to become an emission line \citep{CramGiampapa:1987}. Shallow H$\alpha$ absorption can therefore correspond to either an extremely inactive or intermediately active M dwarf, requiring another activity indicator like \ion{Ca}{2} emission, flare frequency, or UV emission to break the degeneracy.

\subsection{Observations and Reduction} \label{sec:optical_reduction}
We observed 6 northern FUMES targets with the Dual Imaging Spectrograph (DIS) on the Astrophysical Research Consortium (ARC) 3.5 meter telescope at the Apache Point Observatory (APO). DIS is a dual-channel spectrograph with a resolving power $R = \frac{\lambda}{\Delta \lambda}$ ranging from 7000 at the bluest wavelengths to 12000 at the reddest wavelengths for our observing settings. We used the B1200 and R1200 gratings centered on 4400 and $6400 \, \textrm{\AA}$, covering the wavelength ranges 3770 -- 5030 \r{A} and 5880 - 7000 \r{A} for the blue and red arms respectively. We observed 4 southern FUMES targets with the Gemini Multiple Object Spectrograph on the Gemini South telescope (GMOS-S) at the southern site of the Gemini Observatory using the B600 grating centered on either 5200 or $5300 \, \textrm{\AA}$ with a 0.5'' mask. The GMOS-S covered wavelengths from 3640 -- 6780 with the 5200 setting and 3740 -- 6880 with the 5300 setting, with a resolving power of $ 7000 \ (\mathrm{blue} \ \mathrm{end}) < R < 13000 \ (\mathrm{red} \ \mathrm{end}) $ across the observed wavelength regime. All APO/ARC 3.5m DIS observations are listed in Table \ref{tab:observations_apo}, and all Gemini GMOS-S observations are listed in Table \ref{tab:observations_gemini}.

\begin{deluxetable}{ccccc}
\tablecaption{APO/ARC 3.5m DIS/1.5" \label{tab:observations_apo}}
\tablehead{\colhead{UT Date} & \colhead{Star} & \colhead{Initial Airmass} & \colhead{Exposure Duration [s]} & \colhead{$N_{\mathrm{exposures}}$}}
\startdata
2017-05-01 & GJ 410 & 1.03 & 180 & 40 \\
2017-05-01 & Feige 56\tablenotemark{*} & 1.22 & 360 & 1 \\
2017-09-14 & LP 247-13 & 1.13 & 360 & 9 \\
2017-09-14 & G 191B2B\tablenotemark{*} & 1.26 & 360 & 1 \\
2017-09-14 & LP 55-41 & 1.33 & 420 & 4 \\
2017-09-14 & G 249-11 & 1.32 & 480 & 5 \\
2017-09-20 & GJ 49 & 1.16 & 180 & 2 \\
2017-09-20 & GJ 49 & 1.15 & 150 & 21 \\
2017-09-20 & G 191B2B\tablenotemark{*} & 1.26 & 360 & 1 \\
2017-09-20 & GJ 4334 & 1.28 & 360 & 3 \\
2017-09-20 & GJ 4334 & 1.3 & 420 & 13 
\enddata
\tablenotetext{*}{These stars were used as flux standards for their respective observing nights.}
\end{deluxetable}

\begin{deluxetable}{cccccc}
\tablecaption{Gemini South GMOS-S/0.5" B600 Grating \label{tab:observations_gemini}}
\tablehead{\colhead{UT Date} & \colhead{Star} & \colhead{Central Wavelength [nm]} & \colhead{Initial Airmass} & \colhead{Exposure Duration [s]} & \colhead{$N_{\mathrm{exposures}}$}}
\startdata
2017-09-24 & HIP 112312 & 520 & 1.00 & 120 & 6\\
2017-09-24 & HIP 112312 & 530 & 1.00 & 120 & 6\\
2017-09-24 & LTT 9239\tablenotemark{*} & 530 & 1.02 & 90 & 1 \\
2017-09-24 & CD-35 2722 & 520 & 1.17 & 120 & 6\\
2017-09-24 & CD-35 2722 & 530 & 1.13 & 120 & 6\\
2017-11-24 & HIP 23309 & 520 & 1.65 & 120 & 6\\
2017-11-24 & HIP 23309 & 530 & 1.55 & 120 & 6\\
2017-11-24 & EG 21\tablenotemark{*} & 520 & 1.34 & 50 & 1 \\
2017-12-27 & LTT 1020\tablenotemark{*} & 520 & 1.41 & 90 & 1 \\
2017-12-27 & HIP 17695 & 520 & 1.27 & 120 & 6\\
2017-12-27 & HIP 17695 & 530 & 1.33 & 120 & 6
\enddata
\tablenotetext{*}{These stars were used as flux standards for their respective observing nights.}
\end{deluxetable}

We reduced the spectral data from both Gemini and DIS by adapting the pyDIS package \citep{Davenport:2016}. We bias-subtracted, flat-fielded, and then traced the spectrum with a cubic spline. We fit a low-order polynomial to manually identified lines in a CuAr lamp for Gemini data and a HeNeAr lamp for APO data to wavelength calibrate the spectra. We treated the blue and red detector arms of DIS separately as their own datasets. For the Gemini data, we adapted the pyDIS code to match the nature of the raw data; stitching together the multiple chips into one 2D image, filling in the gaps across chips with NaN values, and saving the read noise and gain properties of each chip into a corresponding 2D array instead of the singular float values assumed by the standard pyDIS package. For both the Gemini and DIS data we extracted the spectra using simple aperture box extraction. The code for both reduction pipelines has been included in separate subdirectories of the Zenodo repository associated with this paper\footnote{\zenodorepo \ \citep{zenodo:2013}}. We used the spectral standard stars from the IRAF Spec50Cal catalog (identified in Tables \ref{tab:observations_apo} and \ref{tab:observations_gemini} with an asterisk$^*$) to flux calibrate the spectra, referring to data files included in \texttt{pydis} \citep{Davenport:2016}.

\subsection{Analysis} \label{sec:optical_analysis}
While reducing the optical spectra we noticed a significant flare in the GJ 4334 data and excluded this star from this work's analysis of short-term stochastic variability. For the remaining targets we restricted our analysis to lines that were clearly identifiable in the spectrum, limiting the lowest activity stars to just measurements of \ion{Ca}{2} H and K and/or H$\alpha$.

\subsubsection{Equivalent Widths}
\label{sec:equivalent_widths}
We measured equivalent widths (EWs) for all optical lines following the convention that negative values correspond to emission 

\begin{equation}
    \rm{EW} = \int_{\lambda_{\rm{low}}}^{\lambda_{\rm{upp}}} \left(1 - \frac{F_{\lambda}}{F_{\mathrm{continuum}}}\right)d\lambda \quad \quad \mathrm{Note:} \ \rm{EW} < 0 \Rightarrow \ \mathrm{emission}
\end{equation}

\noindent where the interval $[\lambda_{\rm{low}},\, \lambda_{\rm{upp}}]$ defines a narrow wavelength window centered on the spectral line and $F_{\mathrm{continuum}}$ is the median flux density value evaluated across all datapoints falling within two continuum regions defined by wavelength intervals on either side of the spectral line. The typical windows for each line are listed in Table \ref{tab:ew_windows}, although some exposures or targets required individually shifting or restricting windows to ensure the continuum value matched the base of the emission line.  These changes to the positions of the windows were $< 3$ \r{A} and always chosen such that the median continuum intensity value was at the base of the emission line. Changes to the width of the line windows were more significant and chosen with respect to each target to capture the entirety of the line. The shape of the spectrum beneath the Balmer lines changes significantly across the range of effective temperatures in our sample and any choice of windows that is consistent across all stars would fail to accurately capture the true equivalent widths for some of the sample. All wavelength windows were internally consistent for each star to avoid interfering with the measurement of variability.

We assume that the error on the continuum value is negligible when propagating the errors on the numerically measured equivalent widths because it is averaged over many datapoints. We did not fit the lines using Gaussian profiles and a polynomial continuum for a combination of two reasons. The first is that some of the intermediate activity stars in our sample have Balmer lines that are not well described by a Gaussian profile because the combination of line formation processes yields something close to a flat line. The second is that the bluer Balmer lines are in spectral regions that do not have a clearly identifiable continuum, but rather appear to be a superposition of many molecular absorption features that cannot be easily described by a spline or polynomial function during fitting. Equivalent widths were a more stable measurement across all lines, exposures, and targets and therefore more useful for assessing and comparing variability.

\begin{deluxetable}{ccccc}
\tablecaption{\label{tab:ew_windows} These were the typical wavelength windows used to measure the equivalent widths in both the APO and Gemini data. The windows were selected with reference to Table 2 of \citet{Walkowicz:2009} but not strictly followed. Some exposures required minor modifications to the wavelength windows shown here and the tables associated with all equivalent width measurements are available in the paper's linked data repository.}
\tablehead{\colhead{Spectral Line} & \colhead{Vacuum Wavelength [\r{A}]} &  \colhead{Wavelength Window Width\tablenotemark{*} [\r{A}]} & \colhead{Blue Continuum} & \colhead{Red Continuum}}
\startdata
H$\alpha$ & 6562.79 & 2 -- 14 & 6500 -- 6575 &  6575 -- 6625 \\
H$\beta$ & 4861.35 & 14 & 4840 -- 4850 & 4875 -- 4885 \\
H$\gamma$ & 4340.72 & 10 & 4270 -- 4320 & 4320 -- 4360 \\
H$\delta$ & 4101.7 & 9 & 4060 -- 4080 & 4120 -- 4140 \\
\ion{Ca}{2} H + H$\epsilon$ & 3968.47, 3970.1 & 12 & 3952.8 -- 3956 & 3975.8 -- 3982 \\
\ion{Ca}{2} K & 3933.6 & 8 & 3922.8 -- 3926 & 3943 -- 3946 \\
H$\zeta$ & 3889.1 & 8 & 3850 -- 3880 & 3900 -- 3910 \\
H$\eta$ &  3835.4 & 8 & 3801 -- 3810 & 3845 -- 3854 \\
H$10$ & 3797.7 & 6 & 3779 -- 3785.6 & 3806 -- 3814\\
\enddata
\tablenotetext{*}{Varied depending on the width of the line, typically 14 \AA\ but less for cases of narrow absorption}
\end{deluxetable}

\subsubsection{Optical Emission Line Variability}
\label{sec:optical_variability}
\begin{figure}
    \centering
    \includegraphics[scale=0.5]{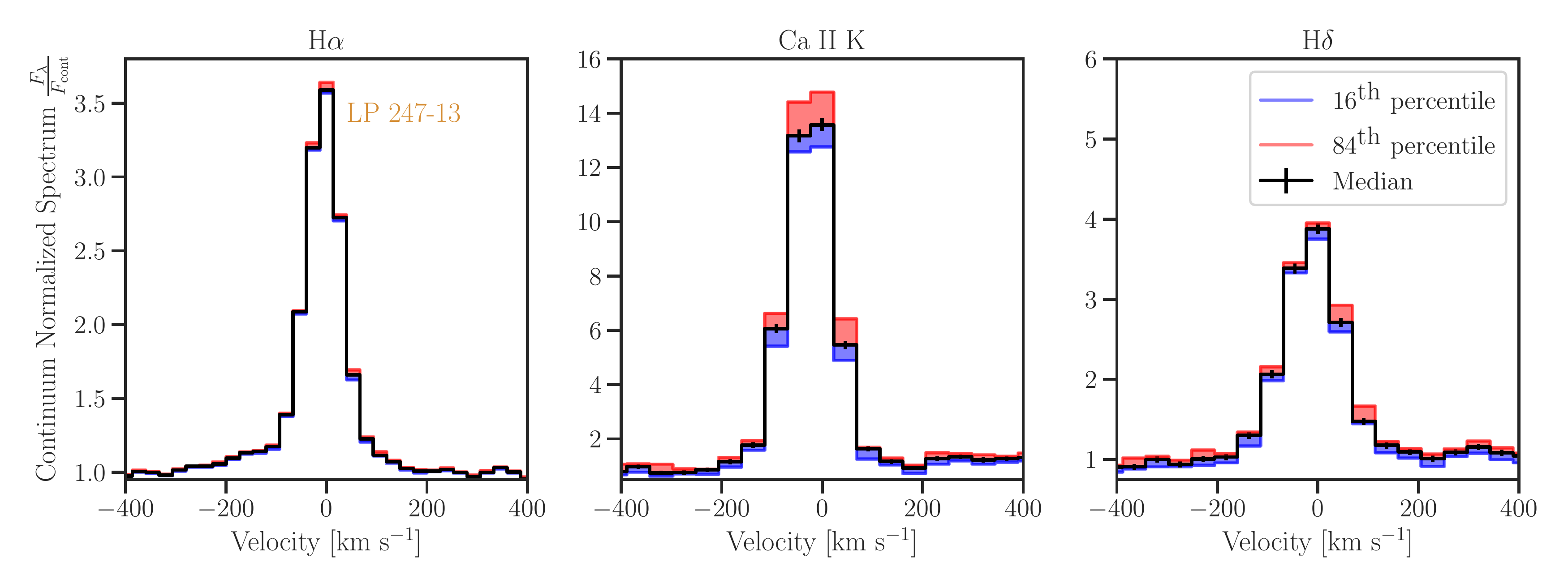}
    \caption{Using LP 247-13 as an example, we demonstrate the variability of three optical emission lines in individual panels; H$\alpha$ on the left, \ion{Ca}{2} K in the middle, and H$\delta$ on the right. We first normalized the spectra from each individual exposure, then resampled them onto a common wavelength grid, and then identified the median spectrum with associated error bars and boundaries for the 16$^{\mathrm{th}}$ and 84$^{\mathrm{th}}$ percentile values of the spectrum in each wavelength bin. The median spectra and their errors are drawn in solid black while the gaps between the median and 16/84$^{\mathrm{th}}$ percentile boundaries are filled in with blue/red respectively. The variability exceeds the error bars in each panel but the magnitude of variation relative to the continuum is higher for \ion{Ca}{2} K and H$\delta$ than for H$\alpha$, following a trend shared by the remainder of the sample where we measured lines other than H$\alpha$.}
    \label{fig:normalized_spectrum_variability}
\end{figure}

With time series measurements of multiple emission lines for the majority of the FUMES sample we were able to assess the minute-to-hour variability of chromospheric emission in these low-mass stars and look for trends between lines and across stars. Figure \ref{fig:normalized_spectrum_variability} shows the variability of LP~247-13's spectrum in three $800 \mathrm{km \, s^{-1}}$ wide regions centered on the continuum-normalized H$\alpha$, \ion{Ca}{2} K, and H$\delta$ emission lines. The black lines and errorbars shows the median value of the continuum normalized flux density across all exposures and the corresponding error. The blue and red shaded regions span the gap between the median and the 16th and 84th percentiles values respectively. Note that this is variability in the emission line relative to the continuum, so this indicates variability in the ratio between chromospheric and photospheric emission that carries forward to the measurements of equivalent widths. Photospheric emission varies by less than a few percent over the course of a full rotation period for low-mass stars like those in our sample \citep{Newton:2016}, and can therefore be neglected in these observations taken over the course of a single night. Normalizing to the continuum removes variations due to observational constraints like seeing, transparency, and instrumental effects.

\begin{figure}
    \centering
    \includegraphics[width=\textwidth]{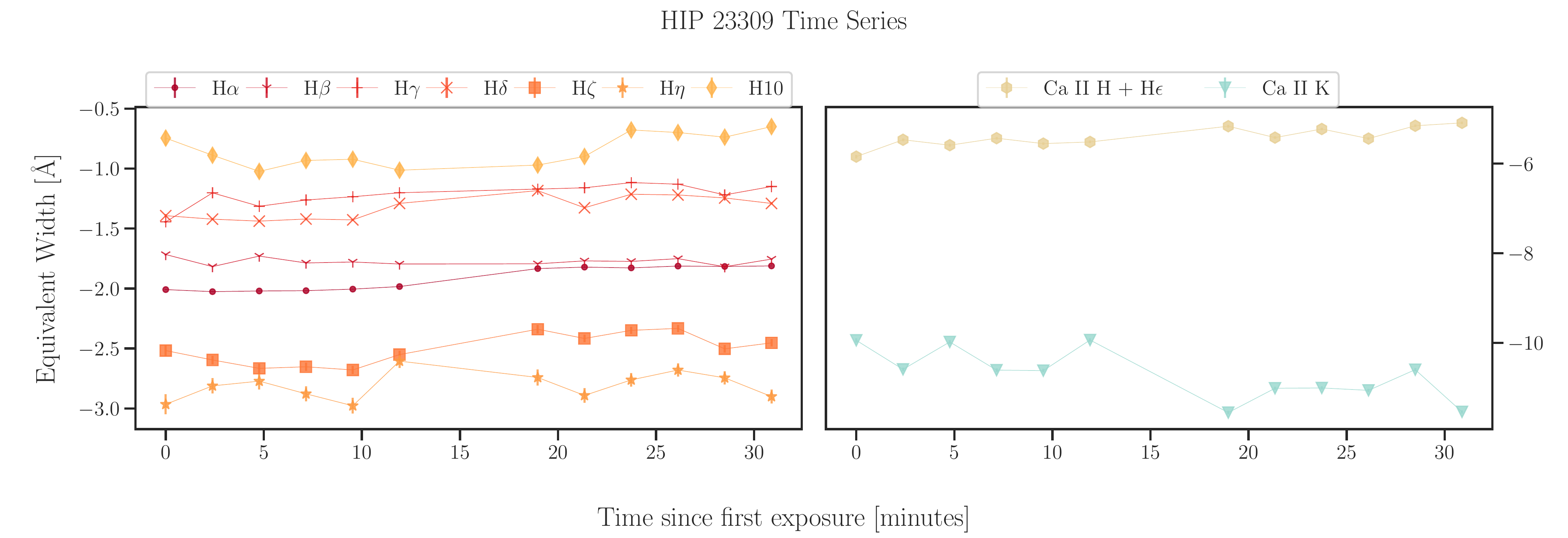}
    \caption{HIP 23309 is one of the more active, UV-bright stars in our sample, and as a result its spectra have signal-to-noise making them ideal for variability analysis on the shortest time scales. Each line's time series is represented by a unique marker shape and color with errorbars, although the errors are minuscule and rarely extend beyond the marker itself. The time series for H$\alpha$'s equivalent width, shown in the darkest red line here, is nearly perfectly flat with no observable long-term trend or short-term variability exceeding the error bars of individual measurements. The lines with the greatest variability are the \ion{Ca}{2} H and K lines which have been plotted in a separate panel for visual clarity, although some of this variability is an offset between grating settings that we attribute to flux calibrating with a standard star spectrum taken with just one of the two settings.}
    \label{fig:lightcurves_single_optical}
\end{figure}

\begin{figure}
    \centering
    \includegraphics[scale=0.5]{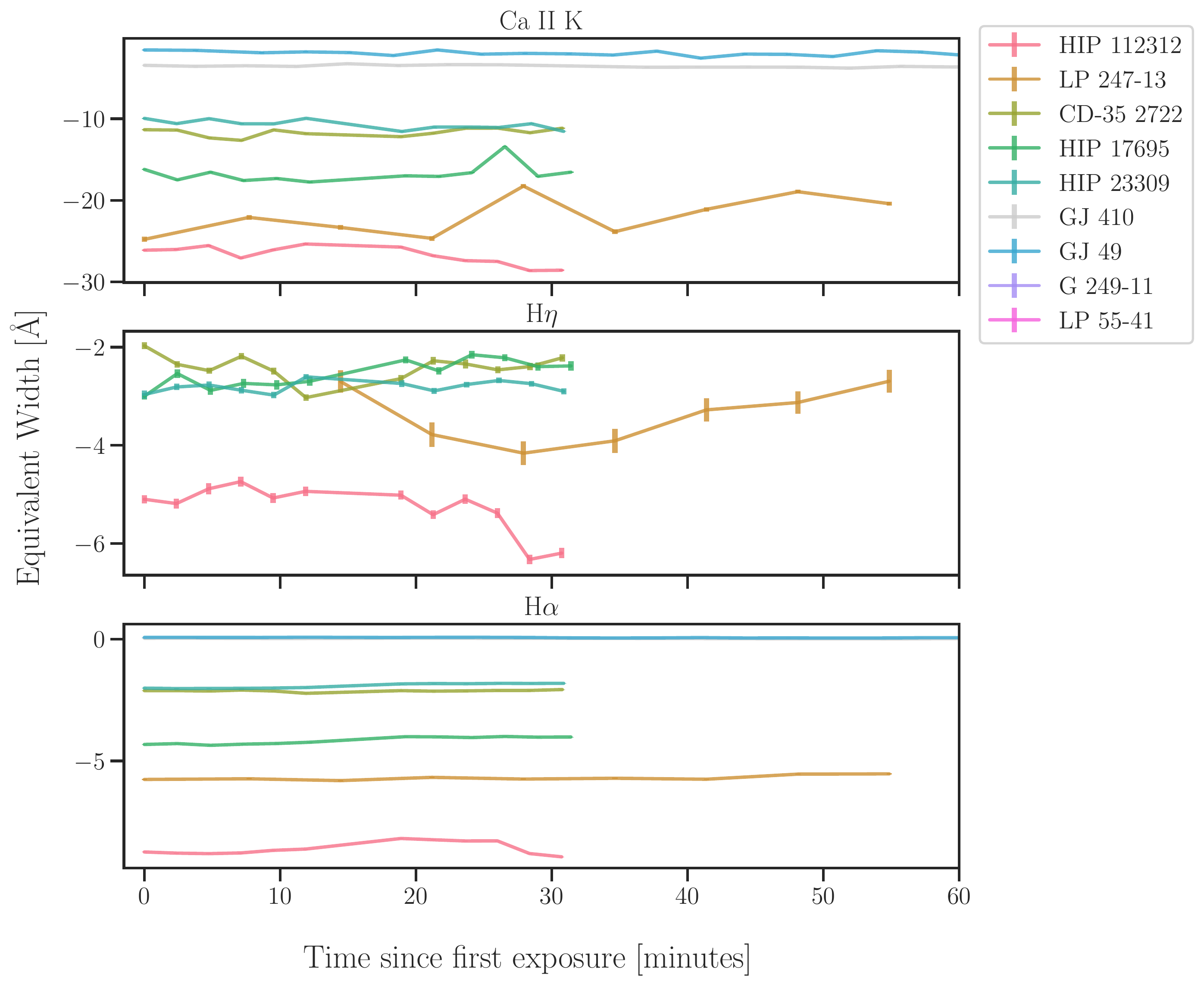}
    \caption{This figure shows the time series measurements for the equivalent widths of \ion{Ca}{2} K (top panel), H$\eta$ (middle panel), and H$\alpha$ (bottom panel) for every star in the sample with measurements of those respective lines. No variability is seen in H$\alpha$  except for one star, HIP 112312. \ion{Ca}{2} and H$\eta$ both exhibit significant variability in all targets. Note that all lightcurves are plotted with errorbars, but that the scale of the error is minuscule, even in Figure \ref{fig:lightcurves_single_optical} which focuses on the optical time series data for just one star.}
    \label{fig:lightcurves_all_optical}
\end{figure}

The line profiles show measurable variation in both amplitude and width over the course of the observation. LP~247-13 was shown as a representative example, and similar variations were observed for the other FUMES stars where we were able to measure lines other than H$\alpha$. Figure \ref{fig:lightcurves_single_optical} shows the time series of equivalent width measurements for all Balmer lines up to H10 and the \ion{Ca}{2} H and K lines for HIP 23309, and Figure \ref{fig:lightcurves_all_optical} shows the time series of equivalent width measurements for three emission lines (H$\alpha$, H$\eta$, and \ion{Ca}{2} K) across the entire sample analyzed in this work. While analyzing the equivalent width time series of Gemini data, we noticed that some of the series showed offsets after the break in the middle corresponding to time spent shifting the central wavelength setting of the grating. We attribute this offset to using only one setting when measuring a standard star spectrum for flux calibration. The offsets vary from $1 \rm{-} 10 \%$ depending on the target and spectral line and can be seen most clearly in the right panel of Figure \ref{fig:lightcurves_single_optical} where the median equivalent width of \ion{Ca}{2} K shifts from $-10$ to $-11$ \r{A} after the gap.

In Figure \ref{fig:normalized_spectrum_variability} the variability of H$\alpha$ is small, only slightly exceeding the errorbars and a small absolute magnitude of variation between the 16$^{\mathrm{th}}$ and 84$^{\mathrm{th}}$ percentile boundaries of the continuum normalized spectrum. This behavior is also apparent in the time series for the H$\alpha$ equivalent width measurements of HIP 23309 shown as dark red in Figure \ref{fig:lightcurves_single_optical}, and again mirrored by nearly every star in the bottom panel of Figure \ref{fig:lightcurves_all_optical}.

\subsubsection{Balmer Decrements}
The Balmer decrement is the flux ratio of each Balmer line (numerically integrated in our equivalent width analysis) relative to either H$\beta$ or H$\gamma$ and can be used to distinguish between the different photoelectric processes forming the Balmer lines \citep{Woolley:1936}. Beyond the limiting cases of pure recombination or collisional excitation, radiative transfer models of stellar atmospheres can be tuned to best match the available data and infer the physical conditions of the photosphere and chromosphere \citep{Houdebine:1994, Allred:2006, Kowalski:2017}. Our measurements of the Balmer series up to H10 provide a valuable dataset to test models of active low-mass stars, bolstered by the measurements of \ion{Ca}{2} H and K which are formed purely by collisional excitation and provide useful points of comparison to the Balmer series \citep{CramGiampapa:1987}. Furthermore, our measurements of the short-term variability of the Balmer series presents opportunities to test models of the inhomogeneity of stellar surfaces and magnetic activity that contribute to this observed variability as perturbations to a steady-state model stellar atmosphere.

\begin{figure}
    \centering
    \includegraphics[scale=0.5]{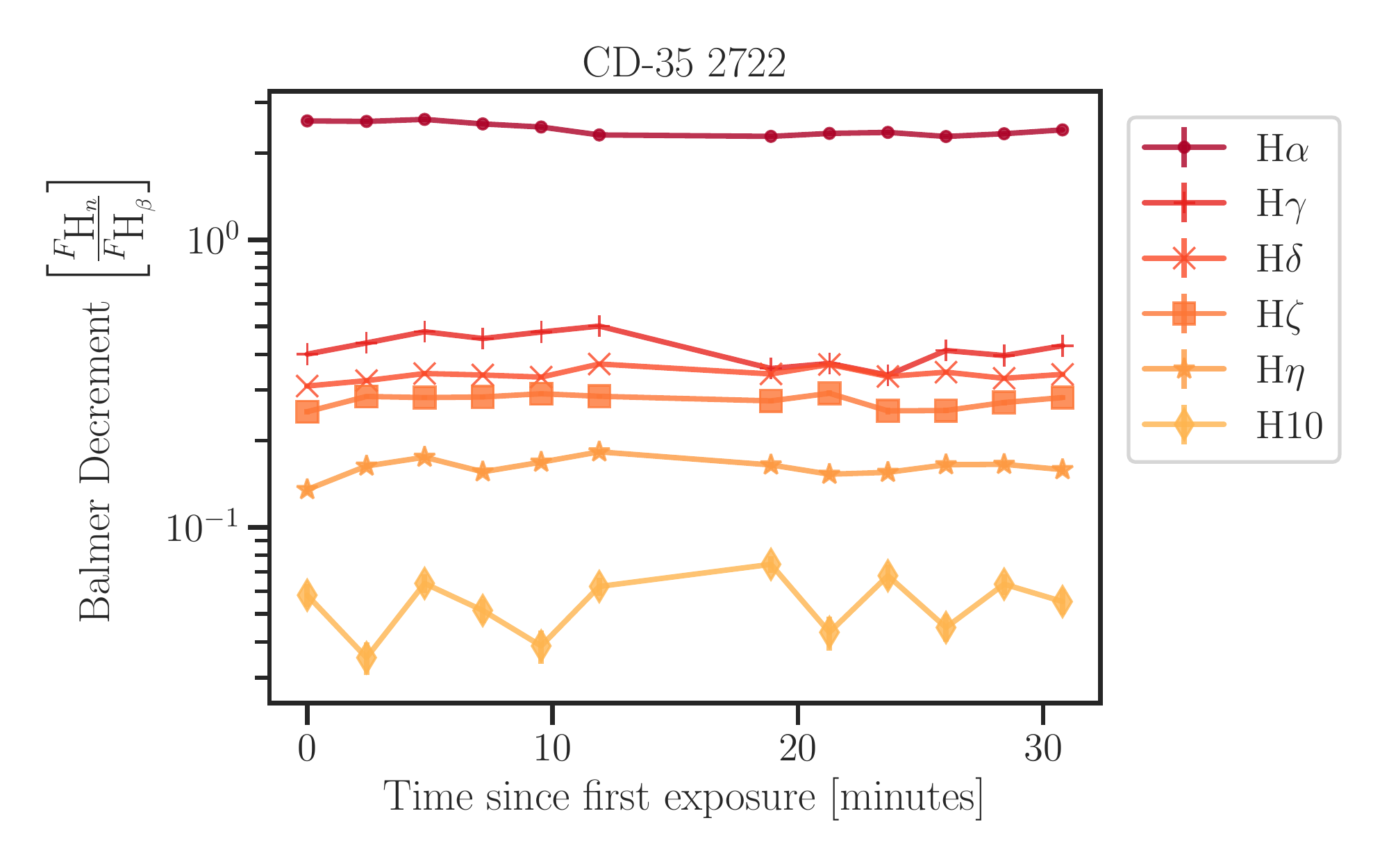}
    \caption{On a logarithmic scale, the time series of CD-35 2722's Balmer decrements show variations exceeding the error bars of individual measurements. A stellar atmosphere model allowed to vary over time with different magnetic heating processes could compare the scale of its decrement variability to this dataset.}
    \label{fig:decrement_series_single}
\end{figure}

\begin{figure}
    \centering
    \includegraphics[scale=0.5]{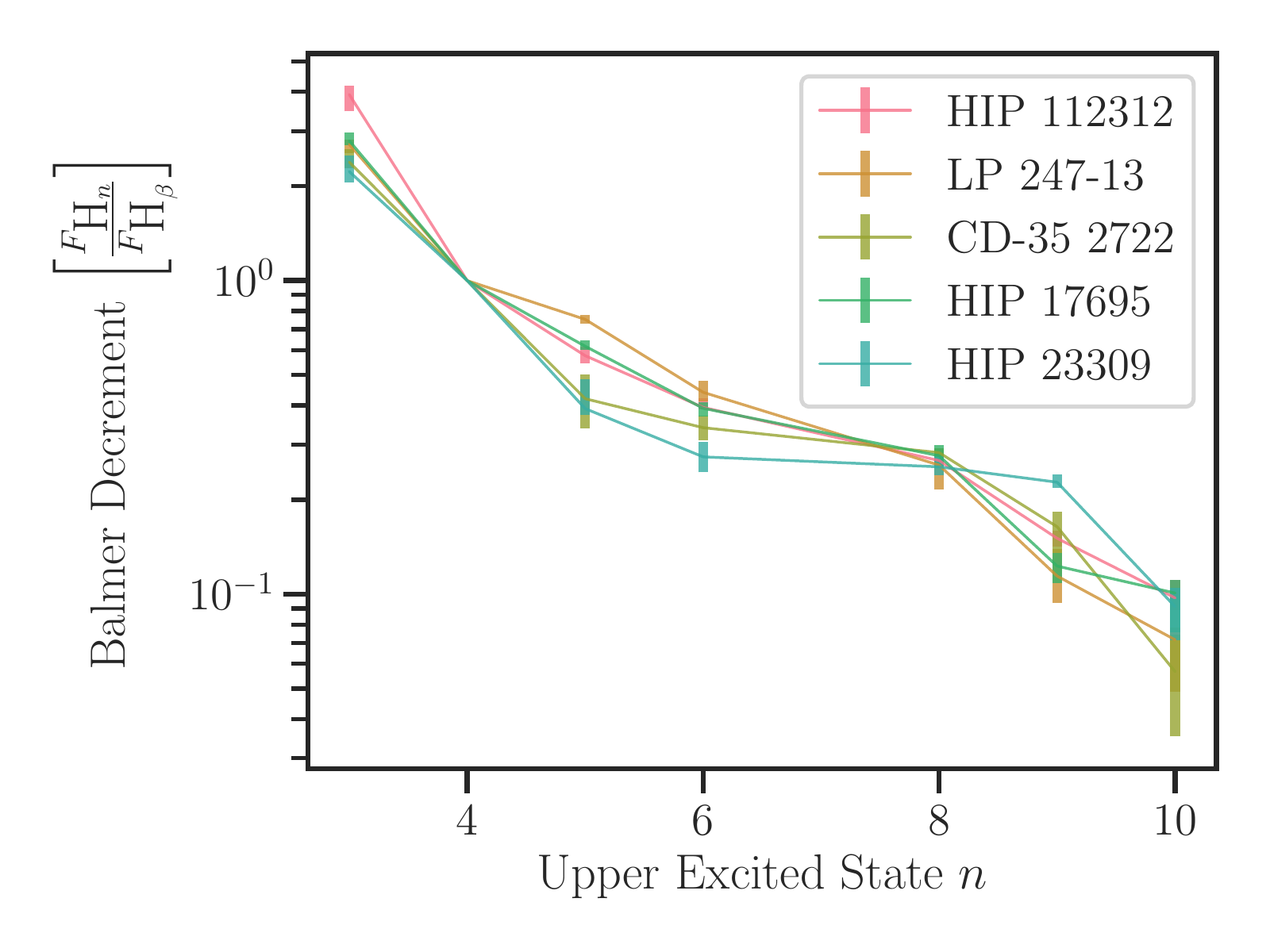}
    \caption{This figure shows the Balmer decrements for all stars in the FUMES sample for which we were able to measure the higher order Balmer lines. The errorbars on the points represent the maximum range of values for the decrement across the time series for that emission line. The values of the Balmer decrements are also recorded in Table \ref{tab:decrements}.}
    \label{fig:decrement_all}
\end{figure}

We compare the ratio of numerically integrated fluxes for the Balmer lines within individual flux-calibrated exposures to the exposure's integrated flux measurement of H$\beta$ to calculate the decrement for each line in each exposure. We do not apply any correction from the photospheric contribution to the H$\alpha$ flux in the decrement analysis to be consistent with the past literature \citep{Hawley:1996, Walkowicz:2009, West:2011}. Figure \ref{fig:decrement_series_single} shows the time series of decrements (except H$\beta$ which is unity by definition) for CD-35 2722 as a representative example while Figure \ref{fig:decrement_all} shows the decrement values of all FUMES stars for which we were able to measure the higher order Balmer lines. Figure \ref{fig:decrement_all} also shows the variability of the decrements with errorbars indicating the maximum range of decrement values. The most variable decrement for most stars is H$\alpha$, which is a consequence of the high variability of H$\beta$ relative to the mostly constant H$\alpha$ (see Figure \ref{fig:lightcurves_all_optical}).

\begin{deluxetable}{cCCCCCC}
\tablehead{\colhead{Star} & \colhead{H$\alpha$} & \colhead{H$\gamma$} & \colhead{H$\delta$} & \colhead{H$\zeta$} & \colhead{H$\eta$} & \colhead{H10}}
\startdata
HIP 112312 & 3.924_{-0.456}^{+0.249} & 0.576_{-0.030}^{+0.039} & 0.394_{-0.006}^{+0.027} & 0.267_{-0.007}^{+0.018} & 0.151_{-0.009}^{+0.008} & 0.097_{-0.008}^{+0.014} \\
LP 247-13 & 2.723_{-0.168}^{+0.097} & 0.752_{-0.025}^{+0.023} & 0.440_{-0.030}^{+0.038} & 0.258_{-0.042}^{+0.035} & 0.114_{-0.020}^{+0.025} & 0.072_{-0.023}^{+0.007} \\
CD-35 2722 & 2.389_{-0.101}^{+0.235} & 0.420_{-0.082}^{+0.081} & 0.340_{-0.030}^{+0.030} & 0.283_{-0.030}^{+0.010} & 0.164_{-0.029}^{+0.019} & 0.057_{-0.021}^{+0.018} \\
HIP 17695 & 2.793_{-0.096}^{+0.174} & 0.618_{-0.014}^{+0.026} & 0.392_{-0.023}^{+0.018} & 0.276_{-0.024}^{+0.023} & 0.123_{-0.014}^{+0.013} & 0.101_{-0.025}^{+0.010} \\
HIP 23309 & 2.228_{-0.169}^{+0.274} & 0.391_{-0.018}^{+0.095} & 0.274_{-0.029}^{+0.031} & 0.255_{-0.015}^{+0.020} & 0.228_{-0.009}^{+0.014} & 0.091_{-0.020}^{+0.014} \\
\enddata
\caption{This table list the median, minimum, and maximum values of the Balmer decrements relative to H$\beta$ for the stars where we were able to measure the higher order lines. The stars are ordered by Rossby number with the lowest/fastest Rossby number on top, corresponding to the most active star.\label{tab:decrements}}
\end{deluxetable}

\section{UV Spectra}\label{sec:uv_analysis}

The brightest far ultraviolet lines are formed in the upper chromosphere (Lyman-$\alpha$, \ion{C}{2}) and transition region (\ion{C}{4}, \ion{N}{5}, \ion{Si}{3}, \ion{Si}{4}). The transition region is a physically thin structure that spans a wide range of temperatures, bridging the $10^4 \,$K chromosphere and $10^6 \,$K corona. FUV emission lines formed here can be used to trace the strength of magnetic heating across this structure, and \citet{Linsky:2020} and \citet{Pineda:2021a} show possible trends in the luminosity of these lines as a function of Rossby number that imply the decline of magnetic heating as stars spin down is first apparent in the corona, then progressively moves inward through the transition region to the chromosphere. We succeed this work to see if the variability of the FUV emission sheds further light on the nature of stellar magnetism and spin-down.

\subsection{HST Data and Analysis}
The FUMES program collected UV spectra using the Space Telescope Imaging Spectrograph (STIS) on \emph{Hubble Space Telescope} (\emph{HST}). Spectra were obtained using the FUV-MAMA detector and the G140L grating for the majority of the targets except HIP 112312 and HIP 17695, for which we used the echelle E140M grating instead. All spectra were taken using the TIME-TAG mode to assess the variability of UV transition region emission lines. See \citet{Pineda:2021a} for a complete description of the observations.

We use \texttt{spectralphoton} \citep{Loyd:2018a, Loyd:2018b} to divide the photon events into time bins of equal duration chosen for each star such that we are able to measure the flux of the \ion{N}{5} doublet to $30\%$ precision in an individual time bin's spectrum. This criterion was chosen because \ion{N}{5} is one of the weaker lines and 30\% seemed like a reasonable compromise between S/N and lightcurve cadence. The durations for each star are listed in Table \ref{tab:uv_time_bins}. We observe two significant flares from GJ 4334 and GJ 410 where the count rate changed by a factor $> 5$. The GJ 4334 flare was observed simultaneously by \emph{HST} and from APO. The analysis of both of these flares is left to future work and our analysis of the short-term stochastic UV variability excludes these high-energy events. 

\begin{figure}
    \centering
    \includegraphics[height=0.5\textheight]{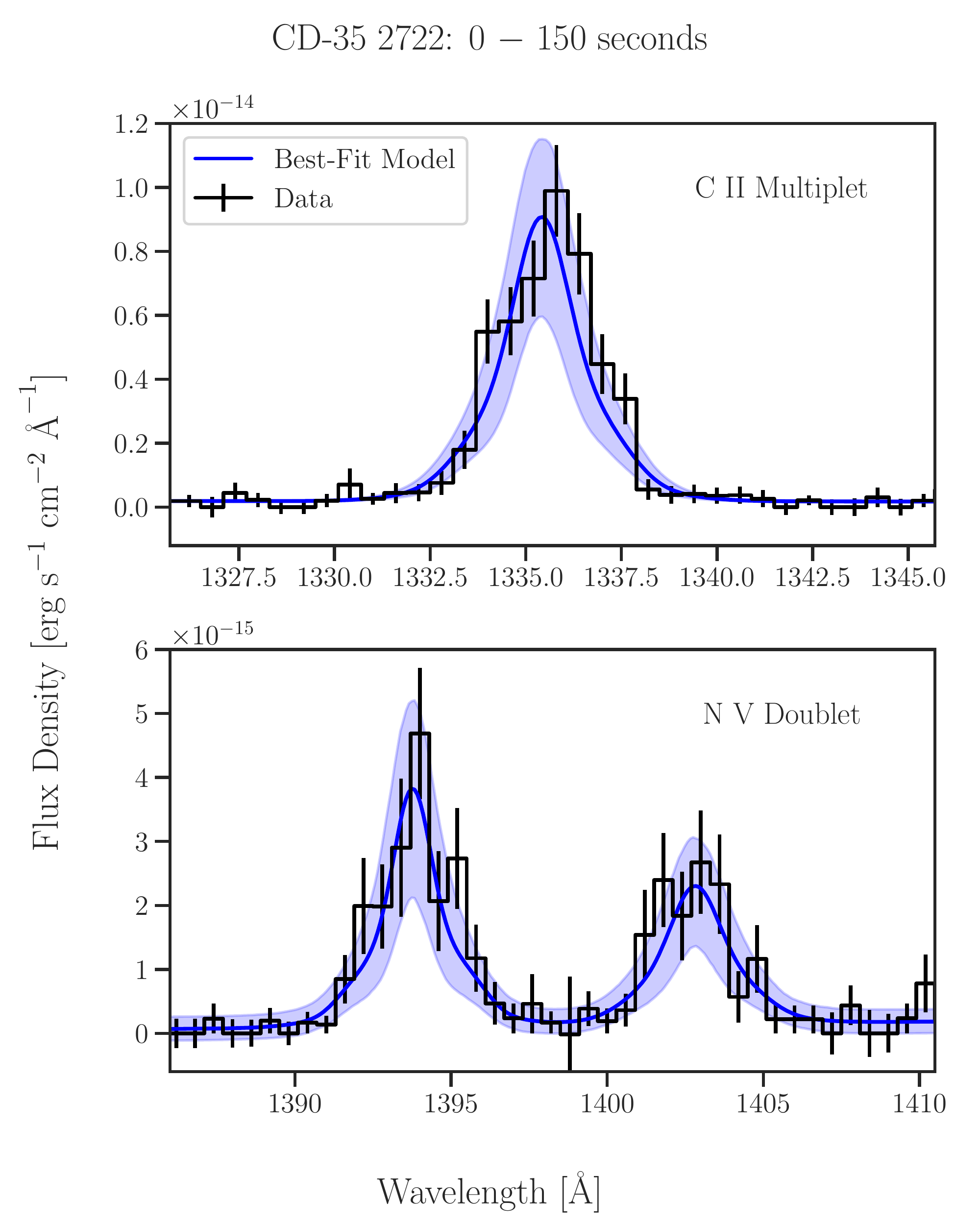}
    \caption{This figure demonstrates the line-fitting process for the first time bin spectrum of CD-35 2722, including photon events that occurred during the first 150 seconds of observation.  The panels show the data in black and best-fit model in blue for the \ion{C}{2} multiplet on top and the \ion{N}{5} doublet on the bottom. The blue swath represents the 1$\sigma$ boundaries of the model fit. Both fits include a sloped continuum, although the slope is very shallow, and the \ion{C}{2} multiplet is fit with a single Gaussian while the \ion{N}{5} doublet is fit using two Gaussians. The models are convolved with the linespread function before evaluating the fit-statistic.}
    \label{fig:linefit_demo}
\end{figure}

With these obvious flares excised, we measured the integrated line fluxes of the \ion{N}{5}, \ion{Si}{4}, \ion{C}{4} doublets, the \ion{C}{2} 1335\,\r{A} multiplet, and the \ion{He}{2} 1640\,\r{A} multiplet for each time bin. We convolve the linespread functions of the observing modes associated with each spectrum during the linefitting process\footnote{\lsfurl}. We fit the lines as Gaussian profiles atop a sloped continuum within windows listed in Table \ref{tab:uv_lines}. The \ion{C}{2} and \ion{He}{2} multiplets were both fit using single Gaussian profiles because we lacked the resolution and S/N to distinguish multiple components. For the \ion{N}{5}, \ion{Si}{4}, and \ion{C}{4} doublets we performed a joint fit of both lines in the doublet using Gaussian profiles for each. We used the \texttt{astropy.models} framework to optimize the fit using the Levenberg-Marquardt non-linear least-squares method \citep{Levenberg:1944, Marquardt:1963}. Figure \ref{fig:linefit_demo} shows an example best-fit to the \ion{C}{2} multiplet and \ion{N}{5} doublet for the first time bin spectrum of CD-35 2722.

\begin{deluxetable}{cc}
\tablecaption{The STIS TIME-TAG mode records individual photon events, enabling studies of variability during an observation. To study the UV variability of these stars we divided each observation into a number of time bins, mimicking exposures, and chose a single duration for each star's time bins based on the S/N, listed here for reference. \label{tab:uv_time_bins}}
\tablehead{\colhead{Name} & \colhead{Time Bin Duration [s]}}
\startdata
HIP 112312 & 500 \\
LP 247-13 & 200 \\
CD-35 2722 & 150 \\
HIP 17695 & 500 \\
HIP 23309 & 100 \\
GJ 49 & 300 \\
\enddata

\end{deluxetable}

\begin{deluxetable}{ccc}
\tablecaption{\label{tab:uv_lines} These ultraviolet lines are formed in the transition regions of low-mass stars \citep{France:2016}. We fit them as Gaussian profiles convolved with the instrumental line-spread function.}
\tablehead{\colhead{Spectral Line} & \colhead{Vacuum Wavelength(s) [\r{A}]} &  \colhead{Wavelength Window [\r{A}]}}
\startdata
\ion{N}{5} doublet & 1238.82, 1242.806 &  1227.6 -- 1254.0 \\
\ion{C}{2} 1335 multiplet & 1334.53, 1335.66, 1335.71  & 1324.0 -- 1347.4 \\
\ion{Si}{4} doublet & 1393.76,  1402.77 & 1381.8 -- 1414.8 \\
\ion{C}{4} doublet & 1548.19, 1550.78 & 1535.5 -- 1558.6 \\
\ion{He}{2} 1640 multiplet & 1640.33, 1640.35, 1640.38, 1640.39, 1640.47, 1640.49, 1640.54 & 1627.9 -- 1653.7 \\
\enddata
\end{deluxetable}

\subsubsection{Ultraviolet Emission Line Variability}
\citet{Pineda:2021a} discusses a long-term trend in the countrate lightcurve for HIP 23309, uniform across the entire FUV spectrum and possibly caused by guiding issues affecting the fraction of starlight that fell on the slit. We follow the method described in \citet{Pineda:2021a} by fitting a third-order polynomial to the trend and dividing it out. The maximum deviation of the countrate from the mean before correcting the trend was $< 20 \%$.

\begin{figure}
    \centering
    \includegraphics[scale=0.5]{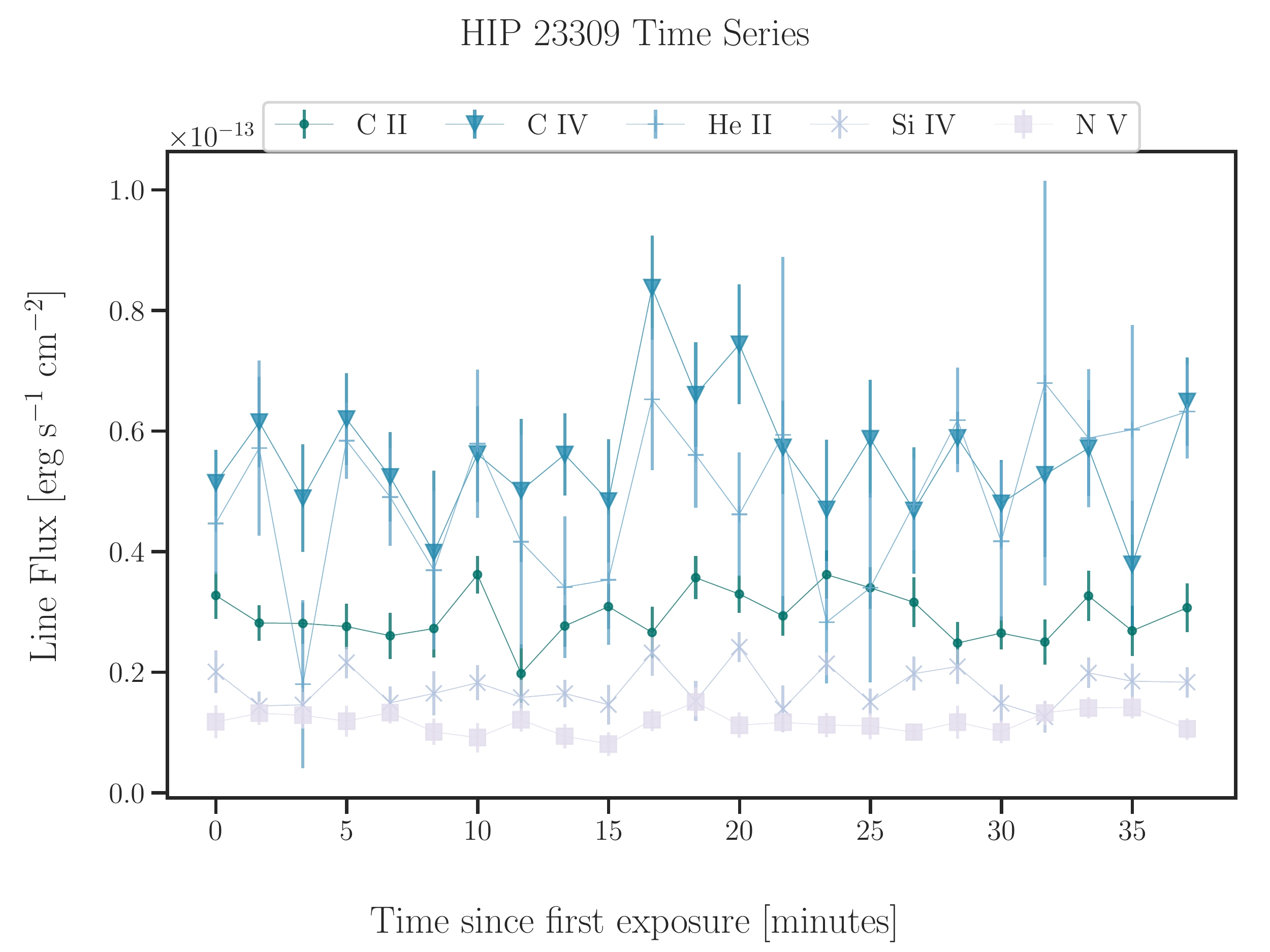}
    \caption{This figure shows the time series of UV line flux measurements for HIP 23309. Unlike Figure \ref{fig:lightcurves_single_optical}, the point-to-point variability is not visibly distinguishable from the photometric uncertainty.}
    \label{fig:lightcurves_single_uv}
\end{figure}

\begin{figure}
    \centering
    \includegraphics[scale=0.5]{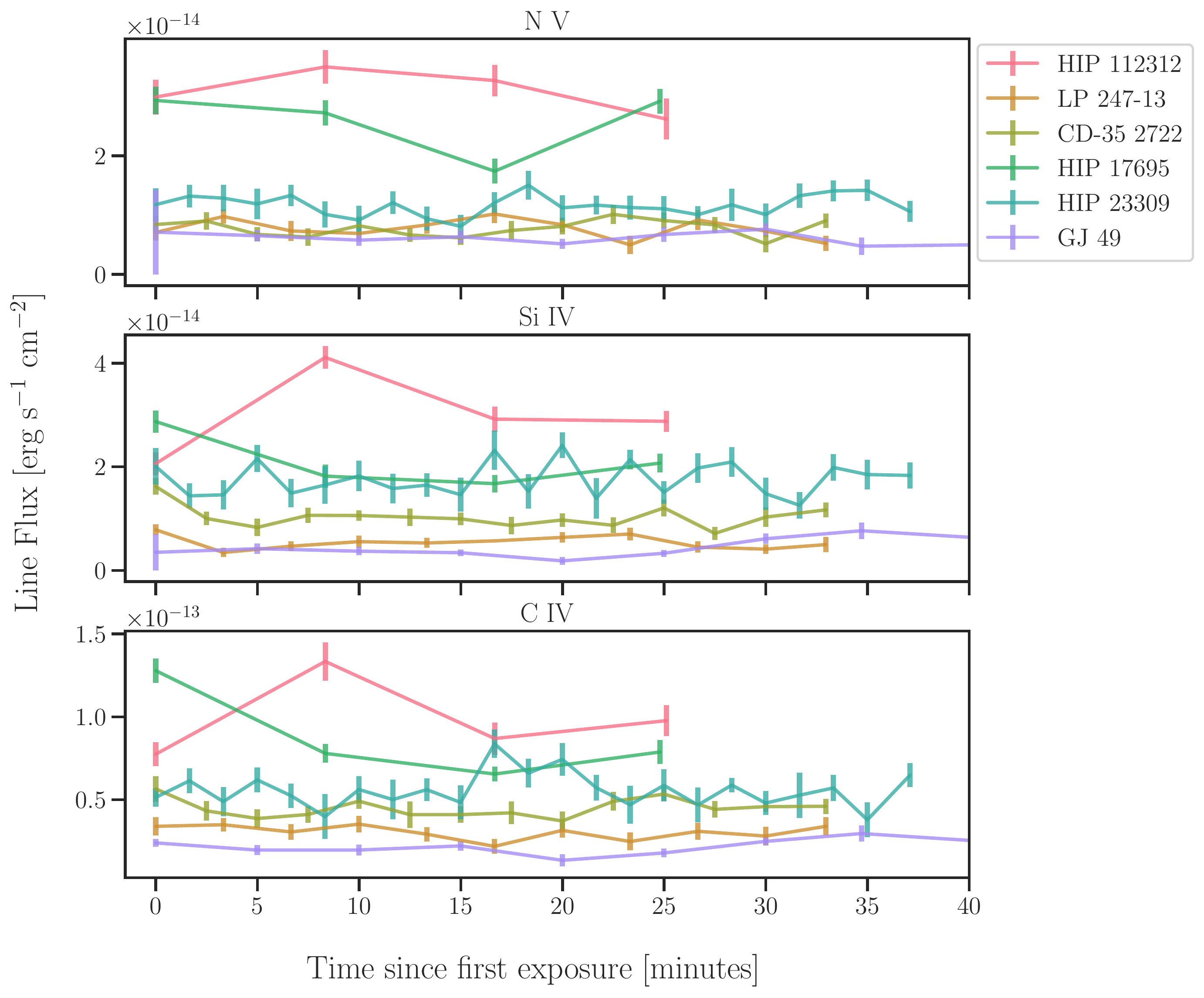}
    \caption{This figure shows the time series measurements of the line fluxes for three UV emission features: \ion{N}{5} in the bottom panel, \ion{C}{4} in the middle panel, and \ion{Si}{4} in the top panel. No clear stochastic or long-term trends are apparent in these time series.}
    \label{fig:lightcurves_all_uv}
\end{figure}

After correcting for these long-term trends and fitting line profiles, we prepare light curves of the integrated flux for all the measured lines, analogous to the optical line light curves (with the caveat that now a higher value indicates more emission). Figure \ref{fig:lightcurves_single_uv} shows all the light curves for emission lines measured in the spectrum of HIP 23309, one of the more active and UV-bright stars in the sample. While there is significant point to point scatter, the changes are roughly the same scale as the error bars associated with each datapoint. Figure \ref{fig:lightcurves_all_uv} shows all measured line light curves for three emission lines: \ion{N}{5} on top, \ion{Si}{4} in the middle, and \ion{C}{4} on the bottom. The light curves for other stars also show scatter consistent with the error bars. To meaningfully compare the variability between lines with statements about statistical significance, we need to quantify the intrinsic variability of line light curves in both the optical and ultraviolet datasets.

\section{Quantifying and Comparing Variability}
\label{sec:variability}
Our data lack the cadence or baseline required to model the variability with a sophisticated functional form or analyze a frequency power spectrum. To quantify the variability we model each light curve as a Gaussian distribution with the mean fixed at the median value of each light curve and a variance that is the quadrature sum of the measurement uncertainty and an intrinsic variability term. Assuming the underlying photospheric continuum has not changed between exposures,

\begin{equation}
    \frac{d \mathrm{EW}}{\mathrm{EW}} \approx - \frac{d F_{\lambda}}{F_{\lambda}} \label{eq:ew}
\end{equation}
meaning that the fractional change in equivalent width corresponds to a fractional change in flux. We analyze both the optical and UV lightcurves in this common framework of dimensionless fractional flux, but also look at the optical lightcurves in absolute equivalent width space and the UV lightcurves in absolute flux space. The fractional flux framework controls for the intrinsic activity of the star and allows comparisons between stars with different activity levels, but also allows comparisons between lines that are intrinsically weaker or formed at different heights and temperatures. Conversely, the absolute flux or equivalent width framework puts the variability in the context of the individual line, distinguishing between a small change in a small quantity from being interpreted as equivalent to a large change in a large quantity. Since this work is focused on chromospheric variability, we use Equation 2 of \citet{Newton:2017} to calculate the photospheric contribution to the H$\alpha$ equivalent width of each star, subtract it off before assessing the variability, and provide the calculated quantity in Table \ref{tab:fumes_sample} for reference. The photospheric contribution to the higher-order Balmer lines is less well-characterized but likely small compared to the emission line strength given that all stars where we measure the higher-order lines are active enough to show them in emission and the photospheric contribution to H$\alpha$ is small.

\begin{figure}
    \centering
    \includegraphics[width=\textwidth]{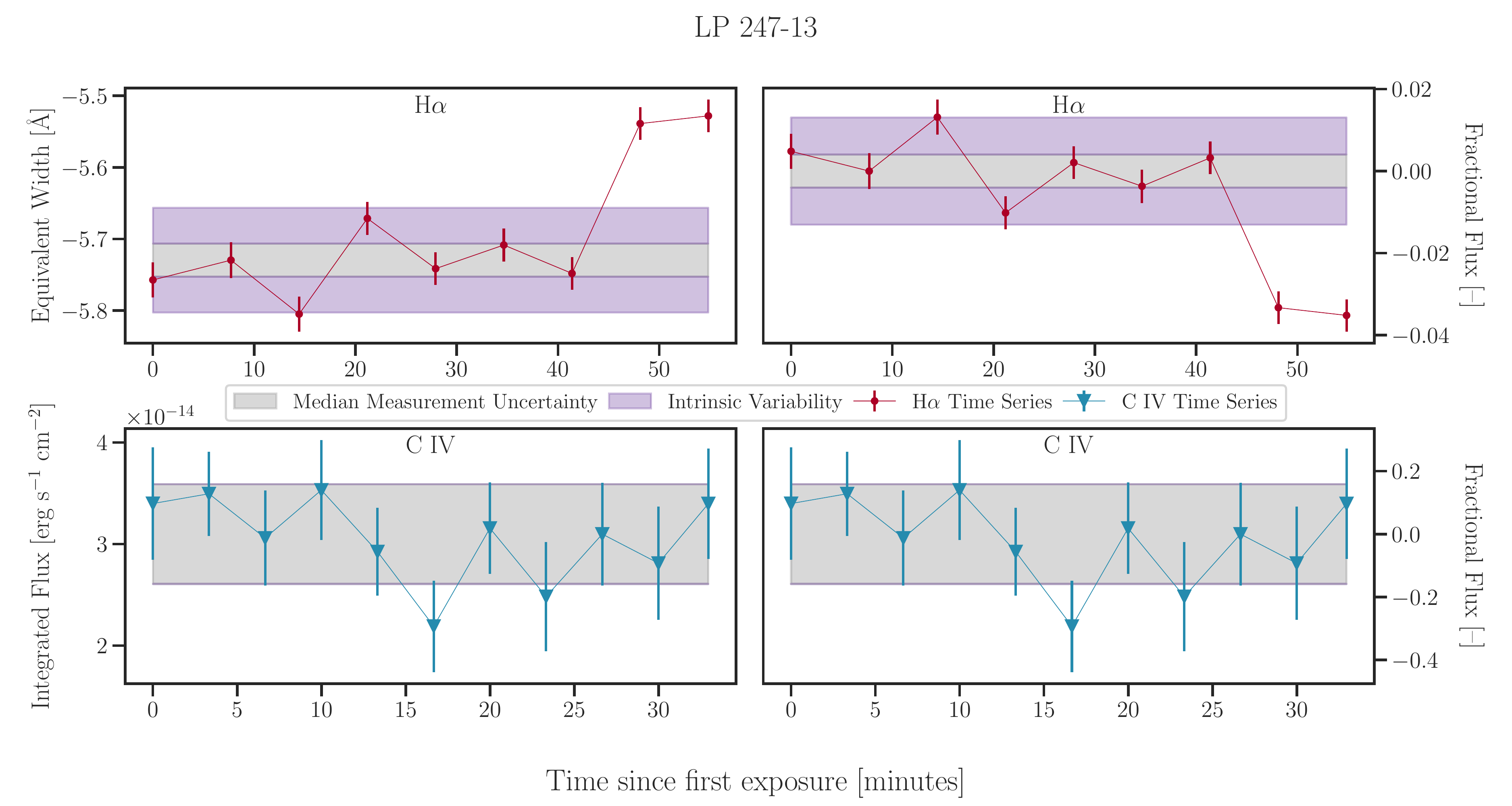}
    \caption{The top panels show the LP 247-13 time series data for H$\alpha$ in absolute equivalent width (left) and dimensionless fractional flux (right) using red dots with error bars. The bottom panels show the LP 247-13 time series data for \ion{C}{4} in a similar fashion, absolute flux on the left and fraction flux on the right. In each panel we use the median measurement uncertainty ($\sigma_{\mathrm{median}}$) of the panel's lightcurve to shade the $\pm1\sigma$ region in gray. Each panel's plotted lightcurve has been fit individually and we use the median of the $s$ parameter posterior distribution to shade the extra contribution in purple such that the boundaries of the shaded region are $\mu \pm \sqrt{\sigma^2_{\mathrm{median}} + s^2}$ where $\mu$ is the median of the lightcurve. The size of the gray swath for the UV data shows how our signal-to-noise prevents us from detecting variability at the same scale as the optical.}
    \label{fig:time_series_demo}
\end{figure}

We use the affine-invariant Markov chain Monte Carlo method implemented by \texttt{emcee} \citep{Foreman-Mackey:2013} to sample the log-likelihood function

\begin{equation}
    \ln \mathcal{L}(y | s) = \sum_{i} \frac{1}{\sqrt{2\pi \left(\sigma_{i}^{2} + s^2\right)}} - \frac{1}{2} \frac{\left(y_i - \mu \right)^2}{\sigma_{i}^{2} + s^2}\label{eq:ew_frac}
\end{equation}
where $\mu$ is the median of the series $y$, which can represent either the absolute equivalent width, absolute flux, or the fractional flux variation. The only parameter being fit is the intrinsic variability $s$ which is either in units of \r{A}, erg s$^{-1}$ cm$^{-2}$, or dimensionless for the absolute equivalent width, absolute flux, and fractional cases respectively. To correct for the offset in Gemini data described earlier in Section \S\ref{sec:optical_reduction} we shift the latter 6 datapoints such that their median matches that of the first 6. We run \texttt{emcee} with 20 walkers for 30000 steps, a factor of $> 10$ times the largest \texttt{emcee}-estimated autocorrelation time of $< 2500$ steps (typical value is $100$), and discard the first 5000 steps. Figure \ref{fig:time_series_demo} demonstrates the four types of fit using LP 247-13 as an example: the optical H$\alpha$ line in the top row, with absolute equivalent width on the left and fractional flux on the right, and the UV \ion{C}{4} line in the bottom row with absolute flux on the left and fractional flux on the right. The gray shaded regions represent the $\pm 1\sigma$ boundaries using the median measurement uncertainty while the purple shaded regions represent the added contribution from the intrinsic variability parameter fit for each panel. In this case, both optical panels show that the intrinsic variability is comparable to the measurement uncertainty while both of the ultraviolet panels' intrinsic variability is a barely visible purple sliver beyond the gray of the measurement uncertainty swath. The comparison between absolute and fractional for both wavelength regimes is effectively just a shift and scaling by the median, although the sign flips for the equivalent width because an increase in flux corresponds to a more negative equivalent width as outlined in Equations \ref{eq:ew} and \ref{eq:ew_frac}.

\begin{figure}
    \centering
    \includegraphics[height=0.4\textheight]{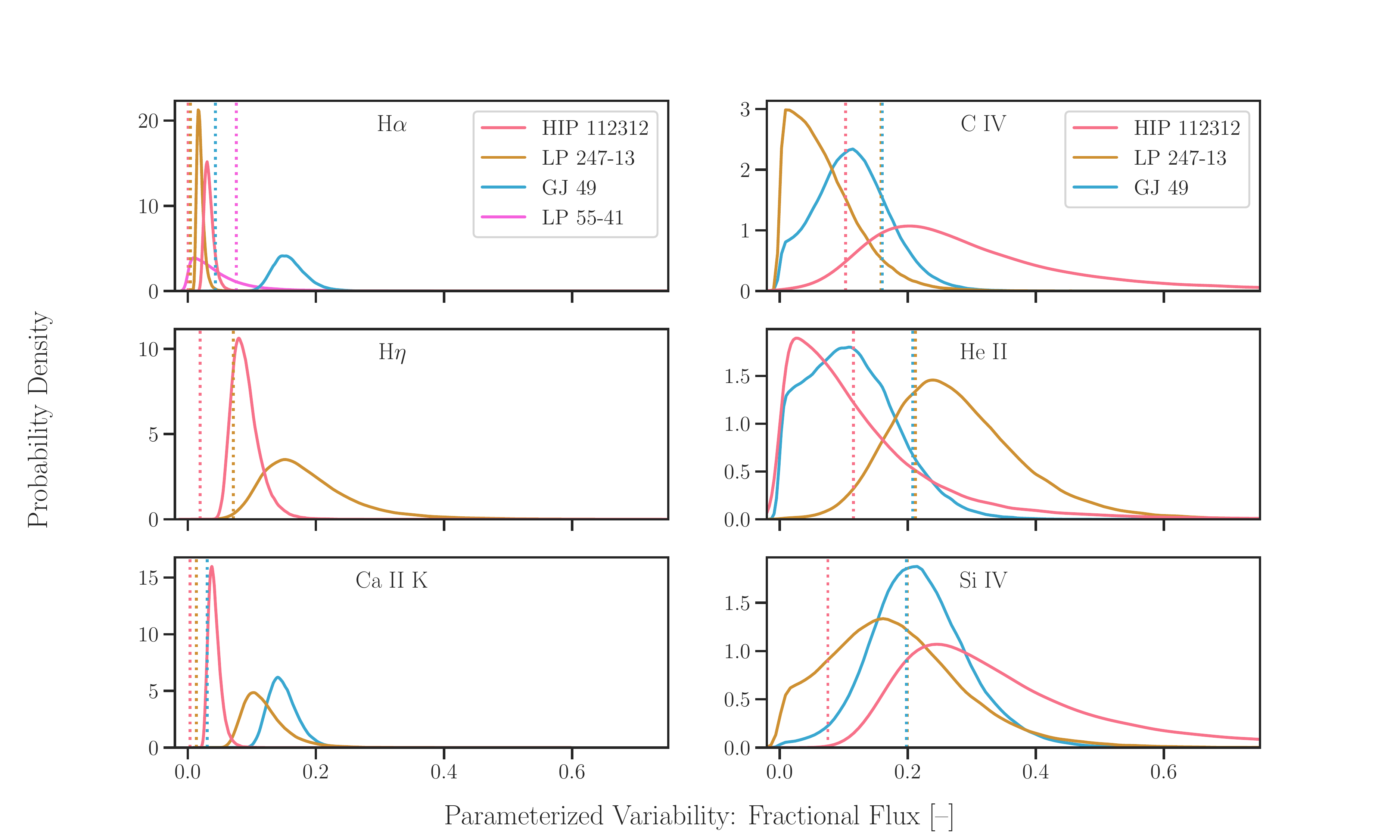}
    \caption{Each panel on the left side shows the posterior distribution for fitting the intrinsic variability in fractional flux for a different optical line: H$\alpha$ on top, H$\eta$ in the middle, and \ion{Ca}{2} K at the bottom. Each panel on the right side shows the posterior distribution for fitting the intrinsic variability in fractional flux for a different ultraviolet emission feature: the \ion{C}{4} doublet on top, \ion{He}{2} 1640 line in the middle, and the \ion{Si}{4} doublet on the bottom. The distributions have been smoothed by a kernel density estimate and are color-coded by star. The median measurement uncertainty for each star's time series is also plotted as a vertical dotted line with the same color-coding although some of these vertical lines are very similar values and appear to overlap. The median measurement uncertainty falls well short of the 16th percentile for most of the optical distributions plotted here, indicating a detection of intrinsic variability that is distinguishable from photometric uncertainty. For most of the ultraviolet distributions, the opposite case is true with the median measurement uncertainty exceeding the 16th percentile and demonstrating that we are unable to detect intrinsic variability that is distinguishable from the noise of the data.}
    \label{fig:multiline_variance}
\end{figure}

Comparing the posterior distribution of the intrinsic variability to the median measurement uncertainty is how we assess the significance of the detected variability. If the median measurement uncertainty exceeds the 16th percentile of the posterior distribution of the parameterized variability $s$, we consider that a non-detection of variability but report the 95th percentile of the posterior distribution as an upper limit. Figure \ref{fig:multiline_variance} compares the measurement uncertainty to the posterior distributions for fractional flux fits to the time series of a few stars for three optical lines on the left side and three ultraviolet emission features on the right. There are only two stars that have significant detections of ultraviolet variability, HIP 112312 and HIP 23309 while most of the optical time series data shows detectable variability. We tabulate the results of all optical fits in Table \ref{tab:optical_variability} and all ultraviolet fits in \ref{tab:uv_variability}. Note that the section of the table for H$\alpha$ measurements have had the photospheric contribution removed and are therefore slightly shifted from the values visible in the lightcurves.

\startlongtable
\begin{deluxetable}{cCCCCCc}
\tablehead{\colhead{Name} & \colhead{Equivalent Width} & \colhead{Absolute intrinsic variability} & \colhead{Fractional intrinsic variability} & \colhead{Number of Measurements} \\
\colhead{--} & \colhead{[\r{A}]} & \colhead{[\r{A}]} & \colhead{--} & \colhead{--}}
\startdata
\cutinhead{Chromospheric H$\alpha$}\\
HIP 112312 & -8.858_{-0.099}^{+0.433} & 0.283_{-0.052}^{+0.075} & 0.032_{-0.006}^{+0.008} & 12 \\
LP 247-13 & -5.982_{-0.025}^{+0.154} & 0.107_{-0.024}^{+0.036} & 0.018_{-0.004}^{+0.006} & 9 \\
CD-35 2722 & -2.410_{-0.018}^{+0.012} & 0.039_{-0.007}^{+0.010} & 0.016_{-0.003}^{+0.004} & 12 \\
HIP 17695 & -4.360_{-0.177}^{+0.126} & 0.071_{-0.013}^{+0.019} & 0.016_{-0.003}^{+0.004} & 12 \\
HIP 23309 & -2.467_{-0.110}^{+0.093} & 0.051_{-0.009}^{+0.014} & 0.020_{-0.004}^{+0.005} & 12 \\
GJ 410 & -0.264_{-0.007}^{+0.006} & 0.008_{-0.001}^{+0.001} & 0.031_{-0.003}^{+0.004} & 40 \\
GJ 49 & -0.210_{-0.016}^{+0.007} & < 0.014 & 0.050_{-0.007}^{+0.009} & 25 \\
G 249-11 & -0.045_{-0.019}^{+0.024} & < 0.058 & < 1.284 & 5 \\
LP 55-41 & -0.070_{-0.001}^{+0.002} & < 0.021 & < 0.297 & 4 \\
\cutinhead{H$\beta$}\\
HIP 112312 & -8.003_{-0.453}^{+0.074} & 0.606_{-0.112}^{+0.159} & 0.075_{-0.014}^{+0.020} & 12 \\
LP 247-13 & -5.771_{-0.440}^{+0.077} & 0.307_{-0.067}^{+0.101} & 0.053_{-0.012}^{+0.018} & 9 \\
CD-35 2722 & -2.361_{-0.081}^{+0.100} & 0.093_{-0.017}^{+0.025} & 0.041_{-0.008}^{+0.011} & 12 \\
HIP 17695 & -4.360_{-0.214}^{+0.103} & 0.073_{-0.014}^{+0.020} & 0.016_{-0.003}^{+0.004} & 12 \\
HIP 23309 & -1.776_{-0.024}^{+0.031} & 0.033_{-0.006}^{+0.009} & 0.018_{-0.003}^{+0.005} & 12 \\
\cutinhead{H$\gamma$}
HIP 112312 & -7.555_{-0.437}^{+0.426} & 0.424_{-0.078}^{+0.113} & 0.054_{-0.010}^{+0.014} & 12 \\
LP 247-13 & -8.452_{-0.528}^{+0.445} & 0.467_{-0.103}^{+0.158} & 0.055_{-0.012}^{+0.019} & 9 \\
CD-35 2722 & -1.570_{-0.239}^{+0.250} & 0.280_{-0.052}^{+0.074} & 0.163_{-0.030}^{+0.043} & 12 \\
HIP 17695 & -4.440_{-0.309}^{+0.184} & 0.128_{-0.024}^{+0.035} & 0.027_{-0.005}^{+0.007} & 12 \\
HIP 23309 & -1.202_{-0.073}^{+0.057} & 0.068_{-0.013}^{+0.018} & 0.054_{-0.010}^{+0.014} & 12 \\
\cutinhead{H$\delta$}
HIP 112312 & -7.291_{-0.412}^{+0.145} & 0.722_{-0.134}^{+0.192} & 0.099_{-0.018}^{+0.026} & 12 \\
LP 247-13 & -5.876_{-0.337}^{+0.347} & 0.329_{-0.078}^{+0.116} & 0.056_{-0.013}^{+0.020} & 9 \\
CD-35 2722 & -2.082_{-0.210}^{+0.076} & 0.143_{-0.027}^{+0.039} & 0.070_{-0.013}^{+0.019} & 12 \\
HIP 17695 & -4.238_{-0.348}^{+0.208} & 0.124_{-0.024}^{+0.034} & 0.027_{-0.005}^{+0.007} & 12 \\
HIP 23309 & -1.309_{-0.114}^{+0.090} & 0.040_{-0.008}^{+0.011} & 0.028_{-0.005}^{+0.008} & 12 \\
\cutinhead{H$\zeta$}
HIP 112312 & -9.480_{-1.060}^{+0.350} & 1.078_{-0.202}^{+0.289} & 0.114_{-0.021}^{+0.030} & 12 \\
LP 247-13 & -7.626_{-0.740}^{+1.105} & 1.006_{-0.258}^{+0.421} & 0.132_{-0.034}^{+0.054} & 7 \\
CD-35 2722 & -3.461_{-0.283}^{+0.271} & 0.273_{-0.054}^{+0.077} & 0.077_{-0.015}^{+0.021} & 12 \\
HIP 17695 & -6.104_{-0.547}^{+0.553} & 0.382_{-0.074}^{+0.106} & 0.058_{-0.011}^{+0.016} & 12 \\
HIP 23309 & -2.510_{-0.146}^{+0.164} & 0.070_{-0.017}^{+0.023} & 0.026_{-0.006}^{+0.009} & 12 \\
\cutinhead{H$\eta$}
HIP 112312 & -5.097_{-0.502}^{+0.173} & 0.442_{-0.086}^{+0.121} & 0.087_{-0.017}^{+0.024} & 12 \\
LP 247-13 & -3.278_{-0.640}^{+0.580} & 0.571_{-0.166}^{+0.258} & 0.173_{-0.051}^{+0.079} & 7 \\
CD-35 2722 & -2.373_{-0.153}^{+0.163} & 0.254_{-0.050}^{+0.072} & 0.108_{-0.022}^{+0.030} & 12 \\
HIP 17695 & -2.511_{-0.285}^{+0.262} & 0.156_{-0.038}^{+0.051} & 0.056_{-0.014}^{+0.018} & 12 \\
HIP 23309 & -2.792_{-0.124}^{+0.065} & 0.074_{-0.027}^{+0.032} & < 0.046 & 12 \\
\cutinhead{H$10$}
HIP 112312 & -3.439_{-0.636}^{+0.148} & 0.449_{-0.089}^{+0.125} & 0.131_{-0.026}^{+0.037} & 12 \\
LP 247-13 & -2.089_{-0.357}^{+0.377} & 0.336_{-0.125}^{+0.180} & 0.161_{-0.060}^{+0.085} & 7 \\
CD-35 2722 & -0.700_{-0.178}^{+0.151} & 0.154_{-0.034}^{+0.047} & 0.238_{-0.053}^{+0.073} & 12 \\
HIP 17695 & -2.125_{-0.212}^{+0.350} & 0.212_{-0.051}^{+0.069} & 0.091_{-0.022}^{+0.030} & 12 \\
HIP 23309 & -0.894_{-0.087}^{+0.200} & 0.133_{-0.028}^{+0.039} & 0.144_{-0.030}^{+0.043} & 12 \\
\cutinhead{\ion{Ca}{2} H + H$\epsilon$}
HIP 112312 & -16.892_{-0.142}^{+0.610} & 1.007_{-0.194}^{+0.281} & 0.060_{-0.011}^{+0.017} & 11 \\
LP 247-13 & -16.281_{-2.240}^{+1.252} & 1.726_{-0.375}^{+0.577} & 0.106_{-0.023}^{+0.035} & 9 \\
CD-35 2722 & -6.508_{-0.372}^{+0.383} & 0.382_{-0.070}^{+0.101} & 0.060_{-0.011}^{+0.016} & 12 \\
HIP 17695 & -10.594_{-0.642}^{+0.745} & 0.476_{-0.089}^{+0.127} & 0.042_{-0.008}^{+0.011} & 12 \\
HIP 23309 & -5.436_{-0.125}^{+0.274} & 0.168_{-0.032}^{+0.046} & 0.030_{-0.006}^{+0.008} & 12 \\
GJ 410 & -4.134_{-0.228}^{+0.284} & 0.238_{-0.025}^{+0.031} & 0.057_{-0.006}^{+0.007} & 39 \\
GJ 49 & 2.634_{-0.124}^{+0.139} & 0.147_{-0.020}^{+0.026} & 0.056_{-0.008}^{+0.010} & 25 \\
\cutinhead{\ion{Ca}{2} K}
HIP 112312 & -26.451_{-1.300}^{+0.763} & 1.050_{-0.195}^{+0.281} & 0.040_{-0.007}^{+0.011} & 12 \\
LP 247-13 & -22.090_{-2.345}^{+2.741} & 2.521_{-0.540}^{+0.818} & 0.114_{-0.024}^{+0.037} & 9 \\
CD-35 2722 & -11.544_{-0.684}^{+0.390} & 0.597_{-0.111}^{+0.159} & 0.052_{-0.010}^{+0.014} & 12 \\
HIP 17695 & -17.021_{-0.486}^{+0.561} & 1.232_{-0.227}^{+0.325} & 0.071_{-0.013}^{+0.019} & 12 \\
HIP 23309 & -10.614_{-0.561}^{+0.643} & 0.510_{-0.095}^{+0.134} & 0.048_{-0.009}^{+0.013} & 12 \\
GJ 410 & -3.646_{-0.116}^{+0.147} & 0.144_{-0.016}^{+0.019} & 0.040_{-0.004}^{+0.005} & 40 \\
GJ 49 & -1.902_{-0.314}^{+0.256} & 0.278_{-0.038}^{+0.047} & 0.146_{-0.020}^{+0.025} & 25 \\
\enddata\
\label{tab:optical_variability}
\tablecaption{This table summarizes all the optical equivalent width measurements, absolute intrinsic variability fits, and fractional intrinsic variability fits with the $+$ and $-$ superscripts delineating the 16th and 84th percentile boundaries of the distributions. In cases where the significant variability is not detected, when the 16th percentile of the posterior is less than the median measurement uncertainty, we report the 95th percentile boundary of the posterior as an upper limit.}
\end{deluxetable}

\startlongtable
\begin{deluxetable}{cCCCc}
\tablehead{\colhead{Name} & \colhead{Integrated Flux} & \colhead{Absolute Intrinsic Variability} & \colhead{Fractional intrinsic variability} & \colhead{Number of Measurements} \\
\colhead{--} & \colhead{[$10^{-15}$ erg s$^{-1}$ cm$^{-2}$]} & \colhead{[$10^{-15}$ erg s$^{-1}$ cm$^{-2}$]} & \colhead{--} & \colhead{--}}
\startdata
\cutinhead{\ion{N}{5} doublet}\\
HIP 112312 & 31.28_{-3.33}^{+2.59} & < 9.235 & < 0.293 & 4 \\
LP 247-13 & 7.32_{-1.10}^{+2.09} & < 2.110 & < 0.286 & 11 \\
CD-35 2722 & 8.14_{-1.83}^{+0.90} & < 1.566 & < 0.192 & 14 \\
HIP 17695 & 28.21_{-6.10}^{+1.05} & < 18.073 & 0.237_{-0.087}^{+0.174} & 4 \\
HIP 23309 & 11.72_{-1.66}^{+1.56} & < 1.452 & < 0.123 & 23 \\
GJ 49 & 5.77_{-0.57}^{+0.87} & < 0.678 & < 0.118 & 19 \\
\cutinhead{\ion{C}{2} multiplet}
HIP 112312 & 27.02_{-4.09}^{+2.72} & < 11.903 & < 0.438 & 4 \\
LP 247-13 & 15.87_{-1.99}^{+1.18} & < 2.107 & < 0.132 & 11 \\
CD-35 2722 & 25.41_{-0.66}^{+0.59} & < 1.604 & < 0.063 & 14 \\
HIP 17695 & 23.37_{-1.21}^{+0.58} & < 5.171 & < 0.220 & 4 \\
HIP 23309 & 28.16_{-1.88}^{+5.30} & < 4.095 & < 0.146 & 23 \\
GJ 49 & 15.78_{-0.92}^{+1.07} & < 1.122 & < 0.071 & 19 \\
\cutinhead{\ion{Si}{4} doublet}
HIP 112312 & 28.99_{-4.45}^{+6.40} & < 24.346 & 0.318_{-0.110}^{+0.224} & 4 \\
LP 247-13 & 5.30_{-0.95}^{+1.34} & < 1.907 & < 0.372 & 11 \\
CD-35 2722 & 10.16_{-1.48}^{+1.44} & < 2.562 & < 0.251 & 14 \\
HIP 17695 & 19.46_{-2.01}^{+5.42} & < 15.783 & 0.295_{-0.111}^{+0.219} & 4 \\
HIP 23309 & 16.49_{-1.88}^{+4.67} & < 3.685 & < 0.224 & 23 \\
GJ 49 & 4.32_{-0.92}^{+1.84} & < 1.529 & < 0.353 & 19 \\
\cutinhead{\ion{C}{4} doublet}
HIP 112312 & 92.32_{-10.32}^{+23.96} & < 69.129 & 0.263_{-0.109}^{+0.208} & 4 \\
LP 247-13 & 30.98_{-4.18}^{+3.39} & < 5.247 & < 0.172 & 11 \\
CD-35 2722 & 43.77_{-2.72}^{+5.37} & < 5.695 & < 0.130 & 14 \\
HIP 17695 & 78.39_{-6.89}^{+25.93} & < 83.096 & 0.406_{-0.140}^{+0.285} & 4 \\
HIP 23309 & 56.11_{-8.59}^{+7.27} & < 8.313 & < 0.149 & 23 \\
GJ 49 & 18.73_{-3.00}^{+3.67} & < 3.983 & < 0.213 & 19 \\
\cutinhead{\ion{He}{2} multiplet}
HIP 112312 & 73.97_{-2.68}^{+12.10} & < 28.200 & < 0.377 & 4 \\
LP 247-13 & 15.47_{-3.41}^{+6.70} & < 7.310 & < 0.470 & 11 \\
CD-35 2722 & 28.54_{-6.02}^{+4.89} & < 6.000 & < 0.210 & 14 \\
HIP 17695 & 40.87_{-1.39}^{+9.06} & < 24.984 & < 0.609 & 4 \\
HIP 23309 & 49.04_{-14.31}^{+11.95} & < 11.789 & < 0.241 & 23 \\
GJ 49 & 10.60_{-2.22}^{+2.64} & < 2.513 & < 0.238 & 19 \\
\enddata\
\label{tab:uv_variability}
\tablecaption{This table summarizes all the ultraviolet integrated flux measurements, absolute intrinsic variability fits, and fractional intrinsic variability fits with the $+$ and $-$ superscripts delineating the 16th and 84th percentile boundaries of the distributions. In cases where the significant variability is not detected, when the 16th percentile of the posterior is less than the median measurement uncertainty, we report the 95th percentile boundary of the posterior as an upper limit.}
\end{deluxetable}

\subsection{GJ 410: An Example of Intermediate Activity}
\begin{figure}
    \centering
    \includegraphics[width=0.4\textwidth]{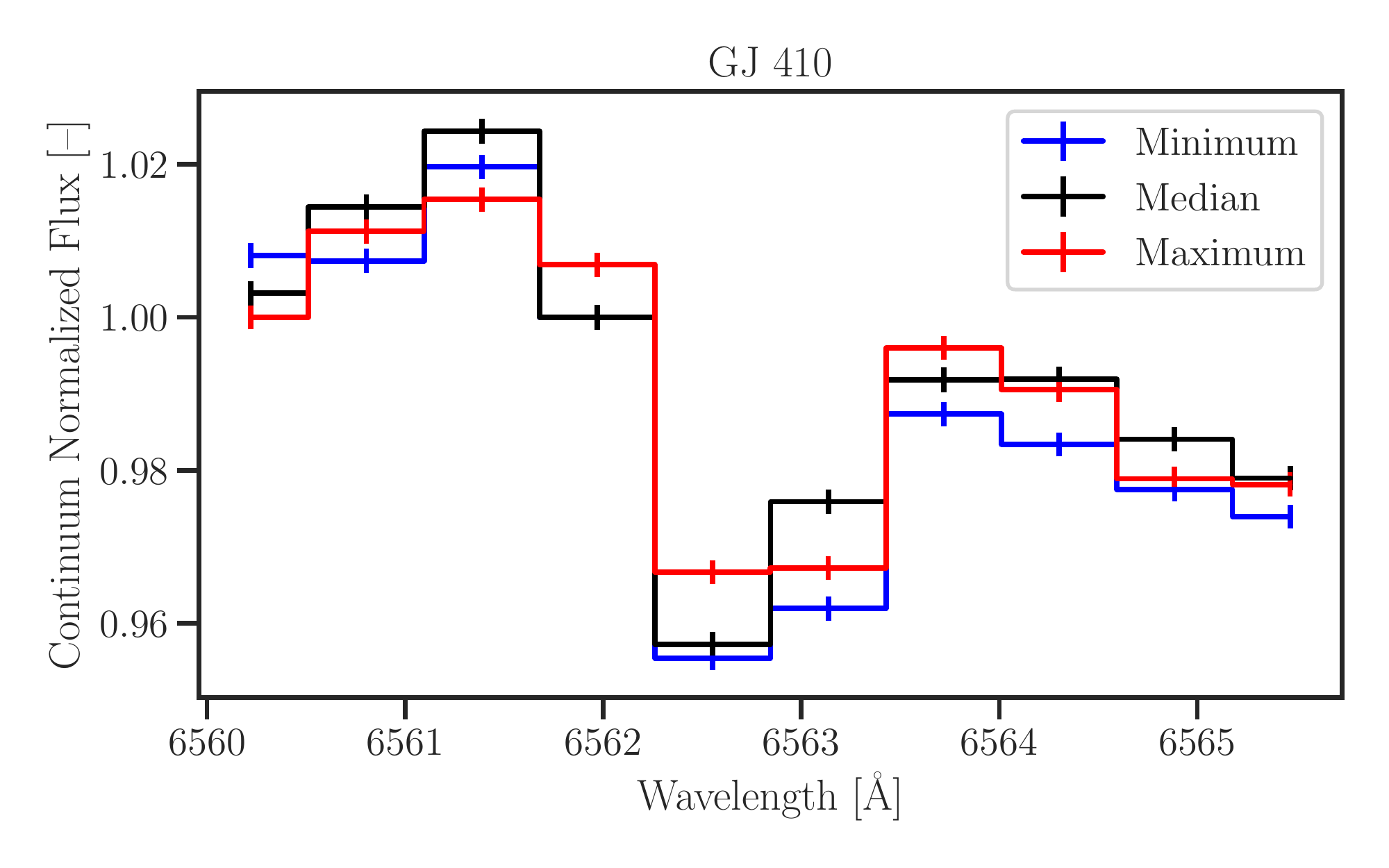}
    \caption{This figure shows the H$\alpha$ spectra for the three exposures of GJ 410 corresponding to the minimum (blue), median (black), and maximum (red) equivalent width measurements in the GJ 410 time series.}
    \label{fig:gj410_fig}
\end{figure}

GJ 410 is at an intermediate activity level where the H$\alpha$ line is very close to being flat, with an equivalent width varying between 0.007 and 0.04 \r{A}. As discussed earlier in Section \S\ref{sec:optical_analysis}, this H$\alpha$ is shallow because it is at an intermediate activity level where the photospheric absorption is almost perfectly filled in by chromospheric emission. These changes are very small in absolute equivalent width but large in relative flux because the average is so close to zero. Figure \ref{fig:gj410_fig} shows the H$\alpha$ variability for this star with continuum normalized spectra. The deepest the line gets is 4\% below the continuum level, with a median equivalent width of 0.025 \r{A}. This skews the apparent variability of GJ 410 and other intermediate activity stars which is why both absolute and fractional variability should be considered in tandem.

If GJ 410 experienced a flare or other brief increase in H$\alpha$ emission, it would go from being barely detectable as either absorption or emission to a definitively detected emission line and likely have a very high ratio between the minimum detected emission line flux to the maximum detected flux. This intermediate activity behavior is also observed by \citet{Kruse:2010} who found that the ratio between maximum and minimum H$\alpha\ $equivalent width was generally higher for stars with intermittently detected H$\alpha$ compared to stars where the H$\alpha$ line was consistently detected.

\subsection{Tentative Trends}
\begin{figure}
    \centering
    \includegraphics[width=\textwidth]{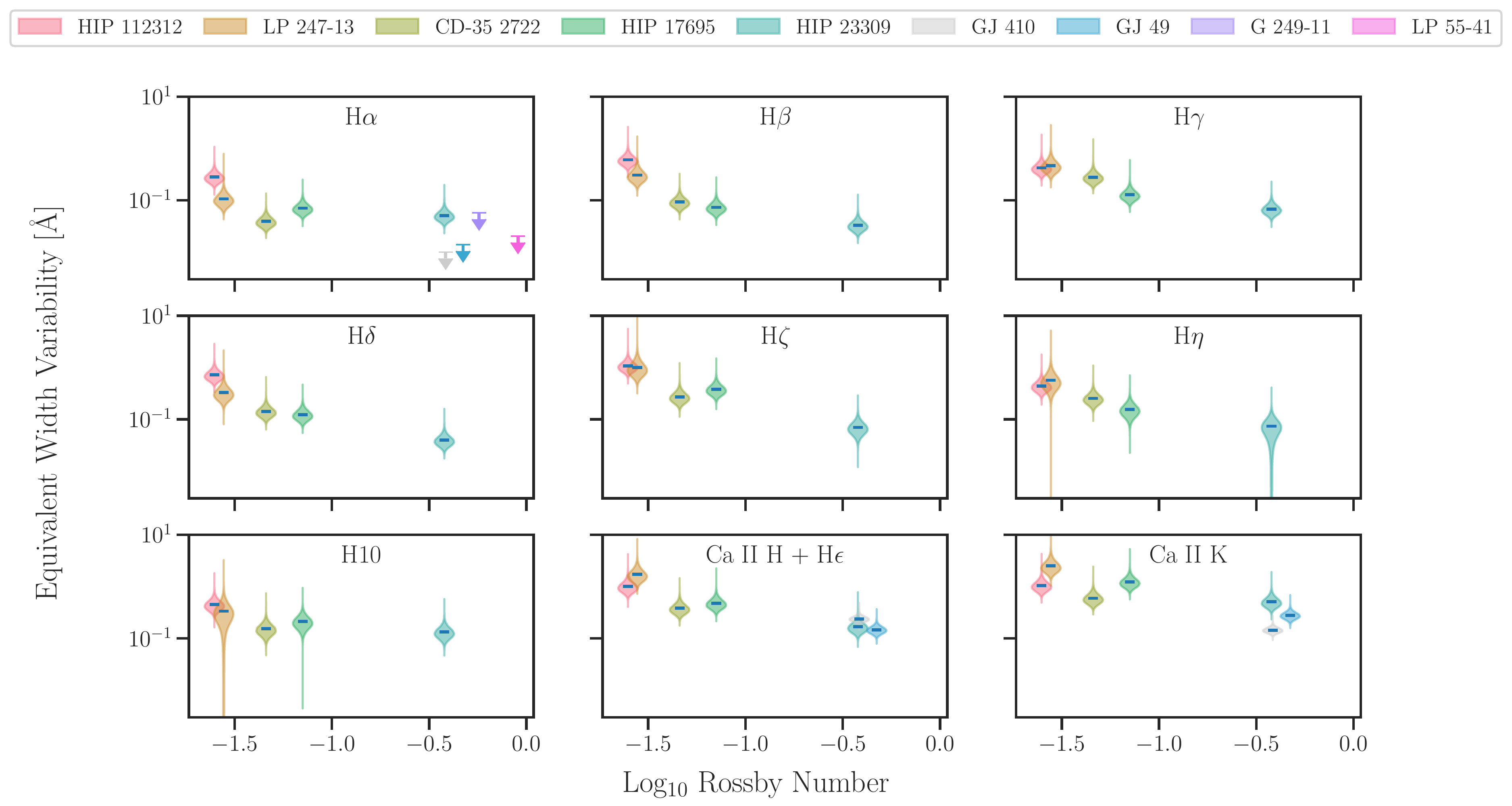}
    \caption{This figure shows the intrinsic variability fit distributions to the optical line equivalent width time series as a function of stellar Rossby number (as calculated in \citealt{Pineda:2021a}). Each distribution is color-coded by star and each panel shows a different line. Smaller Rossby numbers show higher variability and the \ion{Ca}{2} H and K lines show more variability.}
    \label{fig:optical_variability_abs}
\end{figure}

To compare the variability metrics for a single line across the stellar sample we plot them as a function of Rossby number, the ratio between the rotation period and the convective turnover timescale, effectively a mass-normalized rotation metric \citep{Noyes:1984}. The Rossby numbers for the sample, listed in Table \ref{tab:fumes_sample}, were computed in \citet{Pineda:2021a} using the mass determinations from \citet{Pineda:2021b} and the method outlined in \citet{Wright:2018}. Figure \ref{fig:optical_variability_abs} plots the posterior distribution of the equivalent width intrinsic variability fits for all cases where the intrinsic variability was detected. Each panel plots the distributions, color-coded by star, as a function of Rossby number for a different optical line. In this absolute space, lower Rossby numbers (faster rotators) have higher intrinsic variability. Looking at the y-axis values also shows that in this absolute space, the \ion{Ca}{2} H and K lines show more variability than the Balmer series and higher-order Balmer lines show more variability than H$\alpha$.

\begin{figure}
    \centering
    \includegraphics[width=\textwidth]{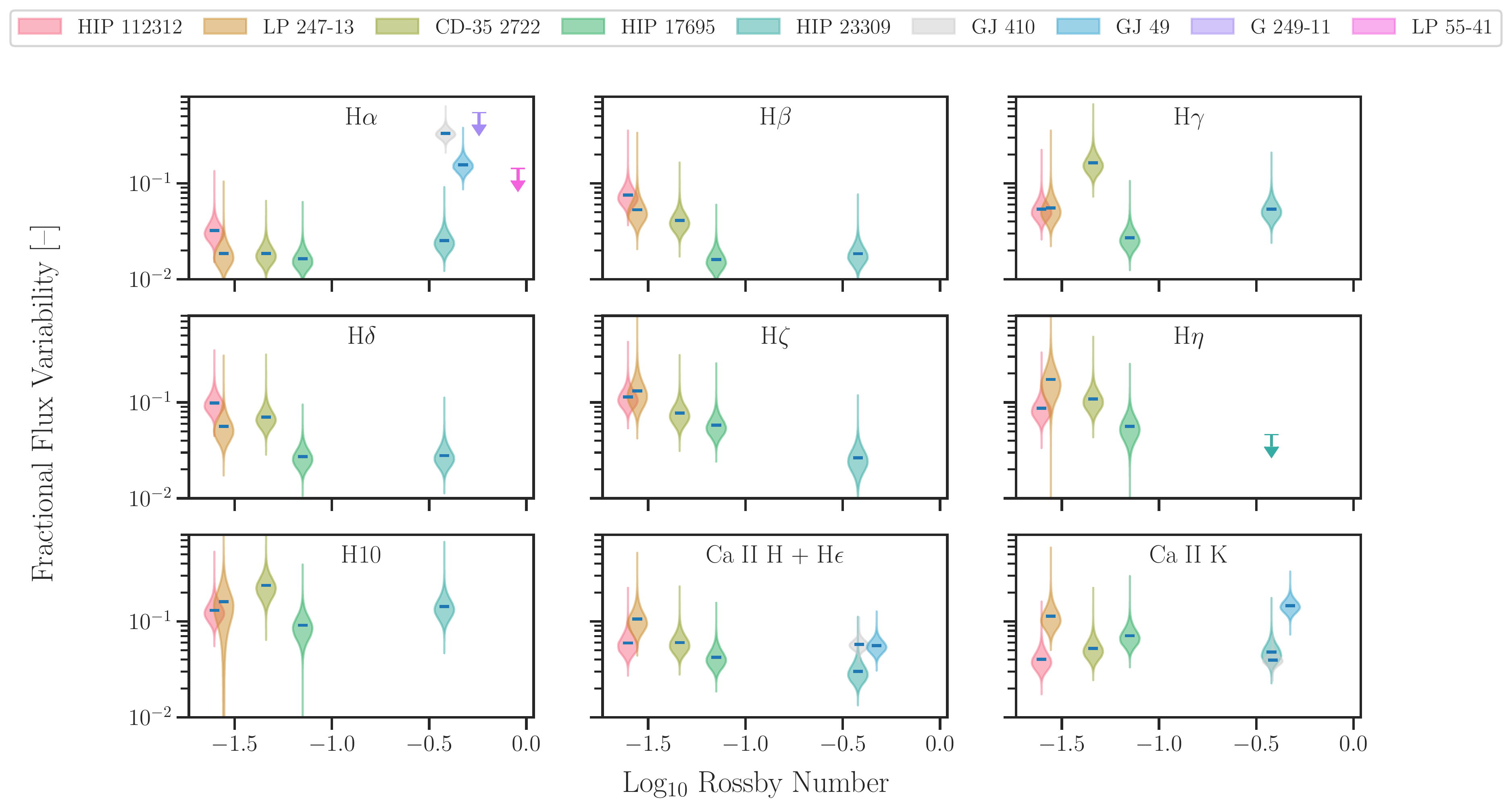}
    \caption{This figure shows the intrinsic variability fit distributions to the optical line fractional flux time series as a function of stellar Rossby number (as calculated in \citealt{Pineda:2021a}). Each distribution is color-coded by star and each panel shows a different line. In this fractional space H$\alpha$ is consistently less variable than the other optical chromospheric lines, but trends with Rossby number are less apparent.}
    \label{fig:optical_variability_frac}
\end{figure}

Figure \ref{fig:optical_variability_frac} is similar to \ref{fig:optical_variability_abs} but plots the intrinsic variability fits to the fractional flux of the optical lines instead. The y-axis is also linear instead of logarithmic. The trends with Rossby number have mostly flattened out indicating that whatever processes contribute to the variability of a line, their behavior as a function of Rossby number is very similar to the behavior of the line's formation as a function of Rossby number. This is conceptually analogous to an observation made by \citet{Loyd:2018b}: flare-frequency distributions of active and inactive M dwarfs differ greatly when flares are characterized in absolute units (flare energy), but appear to coincide when the flares are characterized in relative units (equivalent flare duration, which normalizes by bandpass luminosity). In this relative space, the H$\alpha$ variability is still generally the lowest, but the scatter in \ion{Ca}{2} H and K is broadly consistent with the other Balmer lines.

\begin{figure}
    \centering
    \includegraphics[width=0.956\textwidth]{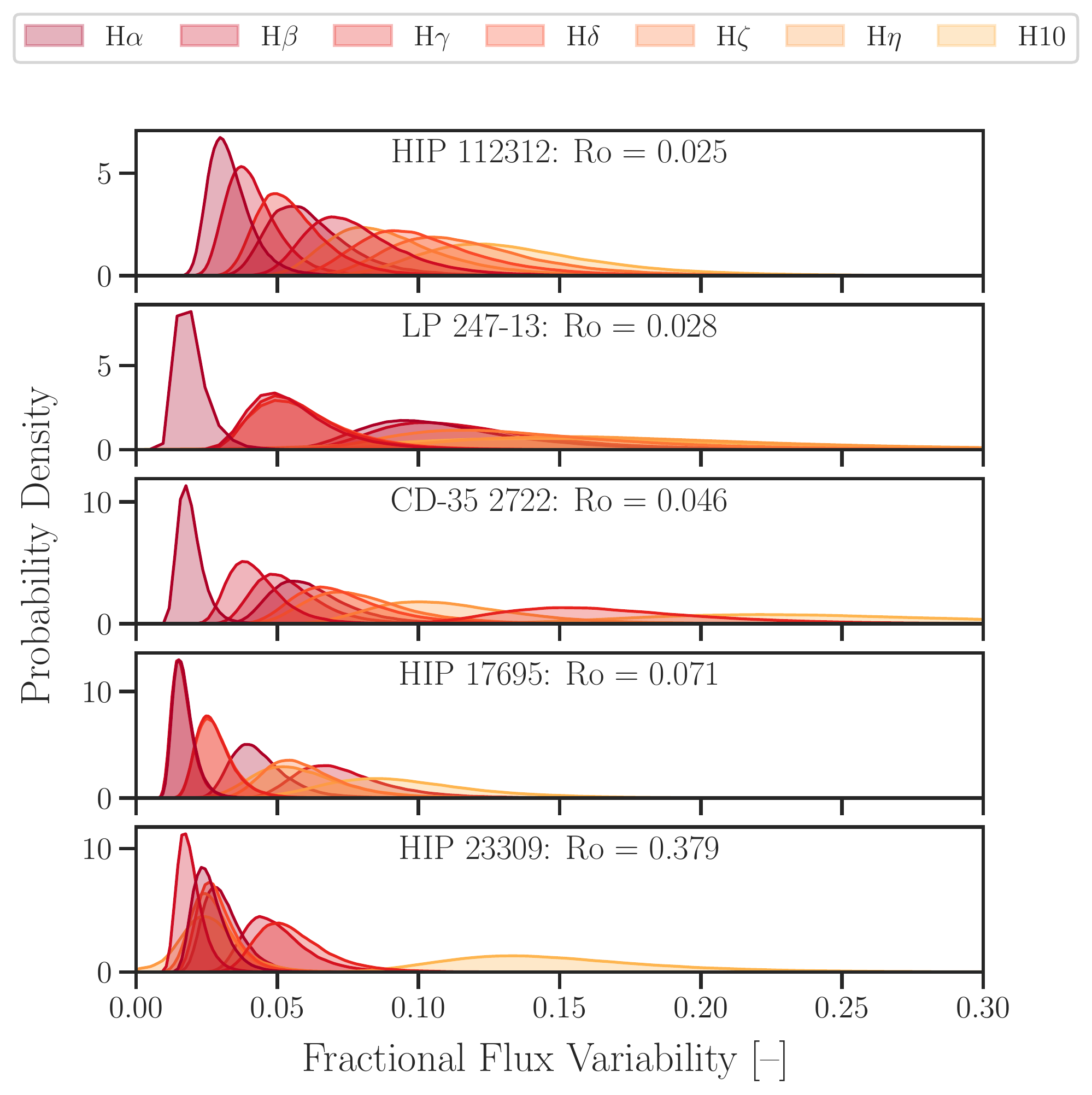}
    \caption{The panels correspond to individual stars, ordered by Rossby number with the lowest Rossby numbers (corresponding to faster rotating and more active stars) on top. Within the panels, the posterior distributions for fitting the intrinsic variability of the fractional flux lightcurves of the Balmer series are plotted with a color-scheme corresponding to the Balmer line order.}
    \label{fig:opt_kde_trend}
\end{figure}

For the few stars active enough that we are able to measure multiple lines in the Balmer series, the parameterized variability in fraction flux shows a tentative trend where higher order Balmer lines have a higher intrinsic variability. The sample is too small to test this rigorously, but Figure \ref{fig:opt_kde_trend} illustrates the pattern. Each panel plots the posterior distributions of fractional flux variability for each Balmer series line, color-coded from red to yellow in increasing order, for one star. The stars are ordered by Rossby number increasing from top to bottom. The fastest rotator, HIP 112312 shows the pattern most clearly where distributions move smoothly from left to right (increasing intrinsic variability) as the color changes from red (H$\alpha$) to yellow (H10). The pattern is less distinct for the other three stars, but is roughly similar.

Previous work by \citet{Lee:2010} has shown that $> 80 \%$ of M3.5V to M8.5V stars are variable (amplitude $> 30\%$) over 1 hour timescales but their sample was focused on active stars with previously known H$\alpha$ emission. \citet{Kruse:2010} looked at a larger sample with a wider range of both spectral type and activity, but had heterogeneous numbers and cadences of measurements. The few stars with $> 5$ measurements were once again biased towards the most active stars. More recently, work by \citet{Medina:2022} has demonstrated variability in the H$\alpha$ line for a 10 star sample of M dwarfs that exceeds our measurements ($50 \textrm{--} 100\%$ equivalent width variability to our $< 30\%$), but their sample includes lower-mass stars than the FUMES sample and their work does not examine trends in the magnitude of line variability as a function of stellar rotation. This work's sample is similarly small, but by selecting a range of Rossby numbers and measuring the higher-order Balmer lines we have found evidence for trends worth examining in more detail for a larger sample of stars.

\subsection{Physically Interpreting Optical Variability}\label{sec:optical_interpretation}

Table \ref{tab:optical_variability} summarizes the findings of our optical variability analysis by listing the Balmer decrements, mean equivalent widths, $s^2_{\small{\mathrm{EW}}}$ values, and number of exposures included in analysis for each emission line. Considering the variability in terms of line formation, if the intrinsic variability in \ion{Ca}{2} H and K is formed by the same mechanisms as the intrinsic variability in the Balmer series, then these mechanisms are related to collisional excitation which dominates the formation of \ion{Ca}{2} H and K \citep{CramGiampapa:1987}. Whatever is causing the variability, the absolute magnitude of the effect is clearly related to the rotation of the star while the relative variability may be constant across Rossby number. More tentatively, higher order Balmer lines are more sensitive to the physical processes corresponding to the variability, which could be due to the higher order lines having lower number densities making small absolute changes more apparent in a relative scale.

 Some possibilities are that these trends may be a function of the total surface area of active regions, the depth of the chromosphere subject to variable energy deposition via microflares or Alfv\'{e}n waves, magnetic field topology, but testing these various possibilities exceeds the scope of this paper. We discuss future avenues for investigation in Section \S\ref{sec:conclusion}.

\subsection{Relative Inability to Detect UV Variability}
Our ability to detect intrinsic variability in the UV was limited by our need to balance time resolution with signal-to-noise ratio. We were generally sensitive to variations beyond 30\% for all the features we measured but do not see evidence of intrinsic stochastic variability exceeding that threshold. This result is largely consistent with the results of \citet{Loyd:2014} which detected stochastic variability in a number of UV targets but few of them showed excess fluctuations $> 30\%$. We were insensitive to UV variability on the scale of what we saw for the majority of the optical line time series.

It is important to note that during the limited time baseline of the FUMES observations, we saw multiple flares, including one on the relatively slow rotator GJ 410, and so these observations join a growing club of UV M dwarf flares \citep{France:2013, France:2020, Loyd:2018a, Froning:2019, Diamond-Lowe:2021}. M dwarfs are therefore unambiguously UV variable, but there is a need to identify whether they show additional stochastic variability that can be distinguished from flares and may have other physical causes. \citet{Loyd:2014} examined this problem for a much larger (and UV brighter) sample, measuring the intrinsic variability left in UV time series data after excising all identifiable flares. For AU Mic, a very active and very nearby M dwarf, \citet{Loyd:2014} found $< 10\%$ variability in the flare-excised data. If UV stochastic variability is caused by magnetic heating processes then one should expect the magnitude of AU Mic's variability to be among the highest in the sample, yet \citet{Loyd:2014} measured higher values of $13.9\%$ and $26.4\%$ for the fainter and less active M dwarfs GJ 832 and Proxima Centauri respectively. But if all or most stochastic variability is caused by flares too small to identify, then AU Mic appears less variable in the flare-excised data because it is much easier to identify the flares. At present this is just supposition, but a carefully planned UV survey with a long baseline could assess whether all M dwarfs show the same amount of stochastic variability after controlling for flare detection completeness.

\section{Conclusion}\label{sec:conclusion}
This paper has collated measurements of the \ion{Ca}{2} H and K and Balmer optical emission lines from H$\alpha$ through H10 for a subset of the FUMES sample, a group of low-mass stars spanning a range of rotation periods observed by \emph{HST} Proposal 14640, and analyzed the variability of both the optical and UV chromospheric emission for this sample. Our measurements of the variability of the optical chromospheric emission indicate stochastic changes to line formation in the chromosphere that correlate to the Rossby number. The H$\alpha$ line is least variable in both an absolute ($< 0.3$ \r{A}) and relative ($< 5\%$, except for GJ 410) sense, while \ion{Ca}{2} H and K vary up to $15\%$ during a single observing night. For the cases where we are able to measure the higher order Balmer lines, their intrinsic variability is $> 5\%$ for all stars except the least active (GJ 410). We see $10 \textrm{--}30\%$ variability in most of the Balmer lines for the lowest Rossby numbers.

Previous studies of H$\alpha$ and \ion{Ca}{2} H and K emission have noted rotational modulation on weekly timescales and activity cycle variations on decadal timescales so rotation-activity relations using these lines have often used the mean equivalent width averaged over archival measurements from multiple nights and attributed the remaining scatter to stellar properties like metallicity \citep{Baliunas:1995, Houdebine:2012, Houdebine:2017}. Our work and \citet{Medina:2022} show that the short-term variation is also a significant contributor that needs to be accounted for and averaging the H$\alpha$ and \ion{Ca}{2} H and K emission over multiple exposures and/or long integration times is necessary to get an accurate measurement of the stellar activity on any given night. This intrinsic variability may also impede studies that try to use these lines to detect star-planet-interactions.

The stochastic optical variability does not coincide with any apparent optical flare continuum emission and no significant stochastic variability is observed in the UV emission lines on similar timescales, only occasional flares where count rate changes significantly ($> $ a factor of 2) across the entire UV spectrum. We are not sensitive to stochastic variability less than $30\%$ in the ultraviolet data which limits our ability to compare the chromospheric variability to the transition region. There are significant discrepancies between the UV and optical flare behavior of low mass stars, namely that UV flares are frequently detected on optically inactive M dwarfs \citep{France:2013, France:2020, Loyd:2018a, Froning:2019, Diamond-Lowe:2021}. If flares on these optically inactive stars skew towards temperatures significantly hotter than 9000 K, then low energy flares may result in observations of optical stochastic variability coincident with UV flares.

We consider two plausible explanations for stochastic variability in the optical line emission that does not observably propagate to the higher layers of the transition region associated with the UV lines we have measured. \citet{Medina:2022} use injection tests to demonstrate that the timescales for flares are consistent with the H$\alpha$ variability they observe. In this scenario a flare produces a cascading beam of non-thermal electrons which collide in the chromosphere, depositing energy which propagates upwards through the transition region to the corona \citep{Hawley:1994}. This could lead to a stratified effect where small (and frequent) flares manifest as small enhancements in the optical emission lines because of the extra collisions, but any flare strong enough to heat the transition region enhances the UV emission lines and continuum dramatically enough to be clearly identifiable as a flare. Another plausible mechanism is Alfv\'{e}n wave heating, where magnetoacoustic oscillations perturb the local pressure, changing the collisional excitation rate, but the strength of these perturbations at different heights would depend on the properties of the stellar magnetic field and atmospheric structure, possibly stratifying to cause more variability in the chromosphere than the transition region \citep{Sakaue:2021}. Determining the relative contribution of each mechanism or identifying other mechanisms is not possible with the dataset we have collected here.

Simultaneous optical and UV monitoring of low mass stars could help determine whether microflaring events are causing the stochastic optical variability. Higher signal-to-noise observations of the UV would enable cross-correlating the variability between the chromosphere and transition region, helping to determine whether the processes involved affect both layers and test for a time lag. A higher cadence and longer baseline survey of the Balmer series for a small sample would enable analysis of the frequency and amplitude of the variability and a comparison between Balmer lines would diagnose the magnitude of energy perturbations. These measurements may be able to disentangle the two heating mechanisms we have described. A theoretical complement to this dataset could be a 3D and time-variable model of the stellar atmosphere where different heating mechanisms can be tuned and trends observed in the simulations can be compared to the available data, applying the techniques of \citet{Kowalski:2017} to stars other than the Sun. A sample structured similarly to FUMES for hotter dwarf stars should also be assembled to determine whether the trends with Rossby number observed here apply to them as well. Following up on existing studies of stochastic variability in one wavelength regime can guide the sample selection, for example getting optical spectroscopic time series data for the \citet{Loyd:2014} UV sample. Magnetic activity is a time-variable phenomenon and time series observations of chromospheric, transition region, and coronal emission of stars across a range of rotation periods and effective temperatures are necessary to understand stellar magnetic structure and evolution.

\begin{acknowledgments}
We thank Alexander Brown and Steven R. Cranmer for conversations that informed and improved this work, as well as the organizers and attendees of the ``Fifty Years of the Skumanich Relations" conference for providing a platform to present and discuss an early version of this work. We would also like to thank the administrative and observing staff at the Apache Point Observatory, the Gemini South Observatory, and the Space Telescope Science Institute. We thank the referee for improving this work with their careful review and comments.

This work is based on observations made with the NASA/ESA Hubble Space Telescope, obtained from the data archive at the Space Telescope Science Institute (STScI) and supported by NASA. STScI is operated by the Association of Universities for Research in Astronomy, Inc. under NASA contract NAS 5-26555. This work is also based on observations obtained at the international Gemini Observatory, a program of NSF’s NOIRLab, which is managed by the Association of Universities for Research in Astronomy (AURA) under a cooperative agreement with the National Science Foundation on behalf of the Gemini Observatory partnership: the National Science Foundation (United States), National Research Council (Canada), Agencia Nacional de Investigaci\'{o}n y Desarrollo (Chile), Ministerio de Ciencia, Tecnolog\'{i}a e Innovaci\'{o}n (Argentina), Minist\'{e}rio da Ci\^{e}ncia, Tecnologia, Inova\c{c}\~{o}es e Comunica\c{c}\~{o}es (Brazil), and Korea Astronomy and Space Science Institute (Republic of Korea). This work is also based on observations  obtained with the Apache Point Observatory 3.5-meter telescope, which is owned and operated by the Astrophysical Research Consortium. We gratefully acknowledge funding for this work through grant numbers HST-GO-14640, HST-GO-15264, and HST-GO-16197 from STScI, and grant number 1945633 from the National Science Foundation NSF/CAREER program. The \emph{Hubble} data used for this work is available at the Mikulski Archive for Space Telescopes: \dataset[10.17909/h3m2-w542]{\doi{10.17909/h3m2-w542}}.

\end{acknowledgments}

%

\vspace{5mm}
\facilities{
    Hubble Space Telescope (Space Telescope Imaging Spectrograph),
    APO/3.5m (Dual Imaging Spectrograph),
    Gemini South Observatory (Gemini Multi-Object Spectrograph)}


\software{
    astropy \citep{astropy:2013,astropy:2018},
    emcee \citep{Foreman-Mackey:2013},
    matplotlib \citep{matplotlib:2007},
    numpy \citep{numpy:2011},
    scipy \citep{scipy:2020},
    seaborn \citep{seaborn:2021},
    pyDIS \citep{Davenport:2016}
    Zenodo \citep{zenodo:2013}
          }



\appendix

\section{Data Product Descriptions}
We provide multiple data products in a Zenodo repository associated with this paper and hosted at \zenodorepo \ \citep{zenodo:2013}. All code associated with the analysis and plots for this paper are included in \texttt{.py} scripts. All raw data for the optical spectra are \texttt{fits} files compressed into a \texttt{gzip} archive while all reduced spectra are in \texttt{fits} files. All tables are provided in the \texttt{astropy} \texttt{ASCII} text \texttt{ecsv} file format.

\begin{itemize}
    \item \textbf{Optical Reduction Code: } The reduction code is divided into two folders, one labelled ``\texttt{pydis}" and another labelled ``\texttt{pygemini}".

    \item \textbf{Equivalent Width Measurements: } The equivalent widths are measured using a combination of two tables for each exposure: one ending with the suffix ``\texttt{ew\_windows.ecsv}" that lists the boundaries of the blue and red continua windows and the continuum flux density value, and another ending with the suffix ``\texttt{ew\_lines.ecsv}" that lists the boundaries of the wavelength window, the integrated line flux with its error, and the equivalent width with its error. These tables are in a subdirectory named \texttt{fit\_tables}. The reduced optical spectra are in a folder labelled ``\texttt{spectra}".

    \item \texttt{spectralPhoton}\textbf{:} The version of \texttt{spectralPhoton} code and the scripts we use to split the \emph{Hubble} \texttt{x1d} spectra are in a directory named ``\texttt{uv}".

    \item \textbf{Line-fitting Tables: } The parameter values and associated errors are recorded in tables structured similarly to the equivalent width tables, ending with suffixes ``\texttt{windows.ecsv}" and ``\texttt{lines.ecsv}".

    \item \textbf{Time Series Tables: } The equivalent widths and integrated line fluxes are collated into time series tables in the subdirectory ``\texttt{time\_series}". Entries with cosmic ray hits in the middle of the line or other spectral defects have been commented out using a \# symbol

    \item \textbf{Posterior Distributions: }All posterior samples and their log-likelihood values are recorded in \texttt{numpy} binary \texttt{.npy} files that can be read using the \texttt{numpy.load()} function. These files are divided into two subdirectories named ``\texttt{abs}"  and ``\texttt{frac}" for the absolute and fractional flux fits respectively.
\end{itemize}

\bibliography{sample631}{}

\begin{thebibliography}{}
\expandafter\ifx\csname natexlab\endcsname\relax\def\natexlab#1{#1}\fi
\providecommand{\url}[1]{\href{#1}{#1}}
\providecommand{\dodoi}[1]{doi:~\href{http://doi.org/#1}{\nolinkurl{#1}}}
\providecommand{\doeprint}[1]{\href{http://ascl.net/#1}{\nolinkurl{http://ascl.net/#1}}}
\providecommand{\doarXiv}[1]{\href{https://arxiv.org/abs/#1}{\nolinkurl{https://arxiv.org/abs/#1}}}

\bibitem[{{Allred} {et~al.}(2006){Allred}, {Hawley}, {Abbett}, \&
  {Carlsson}}]{Allred:2006}
{Allred}, J.~C., {Hawley}, S.~L., {Abbett}, W.~P., \& {Carlsson}, M. 2006,
  \apj, 644, 484, \dodoi{10.1086/503314}

\bibitem[{{Astropy Collaboration} {et~al.}(2013){Astropy Collaboration},
  {Robitaille}, {Tollerud}, {Greenfield}, {Droettboom}, {Bray}, {Aldcroft},
  {Davis}, {Ginsburg}, {Price-Whelan}, {Kerzendorf}, {Conley}, {Crighton},
  {Barbary}, {Muna}, {Ferguson}, {Grollier}, {Parikh}, {Nair}, {Unther},
  {Deil}, {Woillez}, {Conseil}, {Kramer}, {Turner}, {Singer}, {Fox}, {Weaver},
  {Zabalza}, {Edwards}, {Azalee Bostroem}, {Burke}, {Casey}, {Crawford},
  {Dencheva}, {Ely}, {Jenness}, {Labrie}, {Lim}, {Pierfederici}, {Pontzen},
  {Ptak}, {Refsdal}, {Servillat}, \& {Streicher}}]{astropy:2013}
{Astropy Collaboration}, {Robitaille}, T.~P., {Tollerud}, E.~J., {et~al.} 2013,
  \aap, 558, A33, \dodoi{10.1051/0004-6361/201322068}

\bibitem[{{Astropy Collaboration} {et~al.}(2018){Astropy Collaboration},
  {Price-Whelan}, {Sip{\H{o}}cz}, {G{\"u}nther}, {Lim}, {Crawford}, {Conseil},
  {Shupe}, {Craig}, {Dencheva}, {Ginsburg}, {VanderPlas}, {Bradley},
  {P{\'e}rez-Su{\'a}rez}, {de Val-Borro}, {Aldcroft}, {Cruz}, {Robitaille},
  {Tollerud}, {Ardelean}, {Babej}, {Bach}, {Bachetti}, {Bakanov}, {Bamford},
  {Barentsen}, {Barmby}, {Baumbach}, {Berry}, {Biscani}, {Boquien}, {Bostroem},
  {Bouma}, {Brammer}, {Bray}, {Breytenbach}, {Buddelmeijer}, {Burke},
  {Calderone}, {Cano Rodr{\'\i}guez}, {Cara}, {Cardoso}, {Cheedella}, {Copin},
  {Corrales}, {Crichton}, {D'Avella}, {Deil}, {Depagne}, {Dietrich}, {Donath},
  {Droettboom}, {Earl}, {Erben}, {Fabbro}, {Ferreira}, {Finethy}, {Fox},
  {Garrison}, {Gibbons}, {Goldstein}, {Gommers}, {Greco}, {Greenfield},
  {Groener}, {Grollier}, {Hagen}, {Hirst}, {Homeier}, {Horton}, {Hosseinzadeh},
  {Hu}, {Hunkeler}, {Ivezi{\'c}}, {Jain}, {Jenness}, {Kanarek}, {Kendrew},
  {Kern}, {Kerzendorf}, {Khvalko}, {King}, {Kirkby}, {Kulkarni}, {Kumar},
  {Lee}, {Lenz}, {Littlefair}, {Ma}, {Macleod}, {Mastropietro}, {McCully},
  {Montagnac}, {Morris}, {Mueller}, {Mumford}, {Muna}, {Murphy}, {Nelson},
  {Nguyen}, {Ninan}, {N{\"o}the}, {Ogaz}, {Oh}, {Parejko}, {Parley}, {Pascual},
  {Patil}, {Patil}, {Plunkett}, {Prochaska}, {Rastogi}, {Reddy Janga},
  {Sabater}, {Sakurikar}, {Seifert}, {Sherbert}, {Sherwood-Taylor}, {Shih},
  {Sick}, {Silbiger}, {Singanamalla}, {Singer}, {Sladen}, {Sooley},
  {Sornarajah}, {Streicher}, {Teuben}, {Thomas}, {Tremblay}, {Turner},
  {Terr{\'o}n}, {van Kerkwijk}, {de la Vega}, {Watkins}, {Weaver}, {Whitmore},
  {Woillez}, {Zabalza}, \& {Astropy Contributors}}]{astropy:2018}
{Astropy Collaboration}, {Price-Whelan}, A.~M., {Sip{\H{o}}cz}, B.~M., {et~al.}
  2018, \aj, 156, 123, \dodoi{10.3847/1538-3881/aabc4f}

\bibitem[{{Baliunas} {et~al.}(1995){Baliunas}, {Donahue}, {Soon}, {Horne},
  {Frazer}, {Woodard-Eklund}, {Bradford}, {Rao}, {Wilson}, {Zhang}, {Bennett},
  {Briggs}, {Carroll}, {Duncan}, {Figueroa}, {Lanning}, {Misch}, {Mueller},
  {Noyes}, {Poppe}, {Porter}, {Robinson}, {Russell}, {Shelton}, {Soyumer},
  {Vaughan}, \& {Whitney}}]{Baliunas:1995}
{Baliunas}, S.~L., {Donahue}, R.~A., {Soon}, W.~H., {et~al.} 1995, \apj, 438,
  269, \dodoi{10.1086/175072}

\bibitem[{{Basri} \& {Shah}(2020)}]{Basri:2020}
{Basri}, G., \& {Shah}, R. 2020, \apj, 901, 14,
  \dodoi{10.3847/1538-4357/abae5d}

\bibitem[{{Cram} \& {Giampapa}(1987)}]{CramGiampapa:1987}
{Cram}, L.~E., \& {Giampapa}, M.~S. 1987, \apj, 323, 316,
  \dodoi{10.1086/165829}

\bibitem[{{Davenport} {et~al.}(2016){Davenport}, {de Val-Borro}, \&
  {Wilkinson}}]{Davenport:2016}
{Davenport}, J., {de Val-Borro}, M., \& {Wilkinson}, T.~D. 2016, {Pydis:
  Possibly Useful}, v1.1, Zenodo,  Zenodo, \dodoi{10.5281/zenodo.58753}

\bibitem[{{Diamond-Lowe} {et~al.}(2021){Diamond-Lowe}, {Youngblood},
  {Charbonneau}, {King}, {Teal}, {Bastelberger}, {Corrales}, \&
  {Kempton}}]{Diamond-Lowe:2021}
{Diamond-Lowe}, H., {Youngblood}, A., {Charbonneau}, D., {et~al.} 2021, \aj,
  162, 10, \dodoi{10.3847/1538-3881/abfa1c}

\bibitem[{{European Organization For Nuclear Research} \&
  {OpenAIRE}(2013)}]{zenodo:2013}
{European Organization For Nuclear Research}, \& {OpenAIRE}. 2013, Zenodo,
  CERN, \dodoi{10.25495/7GXK-RD71}

\bibitem[{{Foreman-Mackey} {et~al.}(2013){Foreman-Mackey}, {Hogg}, {Lang}, \&
  {Goodman}}]{Foreman-Mackey:2013}
{Foreman-Mackey}, D., {Hogg}, D.~W., {Lang}, D., \& {Goodman}, J. 2013, \pasp,
  125, 306, \dodoi{10.1086/670067}

\bibitem[{{France} {et~al.}(2013){France}, {Froning}, {Linsky}, {Roberge},
  {Stocke}, {Tian}, {Bushinsky}, {D{\'e}sert}, {Mauas}, {Vieytes}, \&
  {Walkowicz}}]{France:2013}
{France}, K., {Froning}, C.~S., {Linsky}, J.~L., {et~al.} 2013, \apj, 763, 149,
  \dodoi{10.1088/0004-637X/763/2/149}

\bibitem[{{France} {et~al.}(2016){France}, {Loyd}, {Youngblood}, {Brown},
  {Schneider}, {Hawley}, {Froning}, {Linsky}, {Roberge}, {Buccino},
  {Davenport}, {Fontenla}, {Kaltenegger}, {Kowalski}, {Mauas}, {Miguel},
  {Redfield}, {Rugheimer}, {Tian}, {Vieytes}, {Walkowicz}, \&
  {Weisenburger}}]{France:2016}
{France}, K., {Loyd}, R.~O.~P., {Youngblood}, A., {et~al.} 2016, \apj, 820, 89,
  \dodoi{10.3847/0004-637X/820/2/89}

\bibitem[{{France} {et~al.}(2020){France}, {Duvvuri}, {Egan}, {Koskinen},
  {Wilson}, {Youngblood}, {Froning}, {Brown}, {Alvarado-G{\'o}mez},
  {Berta-Thompson}, {Drake}, {Garraffo}, {Kaltenegger}, {Kowalski}, {Linsky},
  {Loyd}, {Mauas}, {Miguel}, {Pineda}, {Rugheimer}, {Schneider}, {Tian}, \&
  {Vieytes}}]{France:2020}
{France}, K., {Duvvuri}, G., {Egan}, H., {et~al.} 2020, \aj, 160, 237,
  \dodoi{10.3847/1538-3881/abb465}

\bibitem[{{Froning} {et~al.}(2019){Froning}, {Kowalski}, {France}, {Loyd},
  {Schneider}, {Youngblood}, {Wilson}, {Brown}, {Berta-Thompson}, {Pineda},
  {Linsky}, {Rugheimer}, \& {Miguel}}]{Froning:2019}
{Froning}, C.~S., {Kowalski}, A., {France}, K., {et~al.} 2019, \apjl, 871, L26,
  \dodoi{10.3847/2041-8213/aaffcd}

\bibitem[{{Hawley} \& {Fisher}(1994)}]{Hawley:1994}
{Hawley}, S.~L., \& {Fisher}, G.~H. 1994, \apj, 426, 387,
  \dodoi{10.1086/174075}

\bibitem[{{Hawley} {et~al.}(1996){Hawley}, {Gizis}, \& {Reid}}]{Hawley:1996}
{Hawley}, S.~L., {Gizis}, J.~E., \& {Reid}, I.~N. 1996, \aj, 112, 2799,
  \dodoi{10.1086/118222}

\bibitem[{{Houdebine}(2012)}]{Houdebine:2012}
{Houdebine}, E.~R. 2012, \mnras, 421, 3189,
  \dodoi{10.1111/j.1365-2966.2012.20649.x}

\bibitem[{{Houdebine} \& {Doyle}(1994)}]{Houdebine:1994}
{Houdebine}, E.~R., \& {Doyle}, J.~G. 1994, \aap, 289, 185

\bibitem[{{Houdebine} {et~al.}(2017){Houdebine}, {Mullan}, {Bercu}, {Paletou},
  \& {Gebran}}]{Houdebine:2017}
{Houdebine}, E.~R., {Mullan}, D.~J., {Bercu}, B., {Paletou}, F., \& {Gebran},
  M. 2017, \apj, 837, 96, \dodoi{10.3847/1538-4357/aa5cad}

\bibitem[{{Hunter}(2007)}]{matplotlib:2007}
{Hunter}, J.~D. 2007, Computing in Science and Engineering, 9, 90,
  \dodoi{10.1109/MCSE.2007.55}

\bibitem[{{Kowalski} {et~al.}(2013){Kowalski}, {Hawley}, {Wisniewski}, {Osten},
  {Hilton}, {Holtzman}, {Schmidt}, \& {Davenport}}]{Kowalski:2013}
{Kowalski}, A.~F., {Hawley}, S.~L., {Wisniewski}, J.~P., {et~al.} 2013, \apjs,
  207, 15, \dodoi{10.1088/0067-0049/207/1/15}

\bibitem[{{Kowalski} {et~al.}(2017){Kowalski}, {Allred}, {Uitenbroek},
  {Tremblay}, {Brown}, {Carlsson}, {Osten}, {Wisniewski}, \&
  {Hawley}}]{Kowalski:2017}
{Kowalski}, A.~F., {Allred}, J.~C., {Uitenbroek}, H., {et~al.} 2017, \apj, 837,
  125, \dodoi{10.3847/1538-4357/aa603e}

\bibitem[{{Kruse} {et~al.}(2010){Kruse}, {Berger}, {Knapp}, {Laskar}, {Gunn},
  {Loomis}, {Lupton}, \& {Schlegel}}]{Kruse:2010}
{Kruse}, E.~A., {Berger}, E., {Knapp}, G.~R., {et~al.} 2010, \apj, 722, 1352,
  \dodoi{10.1088/0004-637X/722/2/1352}

\bibitem[{{Lee} {et~al.}(2010){Lee}, {Berger}, \& {Knapp}}]{Lee:2010}
{Lee}, K.-G., {Berger}, E., \& {Knapp}, G.~R. 2010, \apj, 708, 1482,
  \dodoi{10.1088/0004-637X/708/2/1482}

\bibitem[{Levenberg(1944)}]{Levenberg:1944}
Levenberg, K. 1944, Quaterly Journal on Applied Mathematics, 164

\bibitem[{{Linsky}(2017)}]{Linsky:2017}
{Linsky}, J.~L. 2017, \araa, 55, 159,
  \dodoi{10.1146/annurev-astro-091916-055327}

\bibitem[{{Linsky} {et~al.}(2020){Linsky}, {Wood}, {Youngblood}, {Brown},
  {Froning}, {France}, {Buccino}, {Cranmer}, {Mauas}, {Miguel}, {Pineda},
  {Rugheimer}, {Vieytes}, {Wheatley}, \& {Wilson}}]{Linsky:2020}
{Linsky}, J.~L., {Wood}, B.~E., {Youngblood}, A., {et~al.} 2020, \apj, 902, 3,
  \dodoi{10.3847/1538-4357/abb36f}

\bibitem[{{Llama} \& {Shkolnik}(2015)}]{Llama:2015}
{Llama}, J., \& {Shkolnik}, E.~L. 2015, \apj, 802, 41,
  \dodoi{10.1088/0004-637X/802/1/41}

\bibitem[{{Lockyer}(1868)}]{Lockyer:1868}
{Lockyer}, J.~N. 1868, Proceedings of the Royal Society of London Series I, 17,
  131

\bibitem[{{Loyd} \& {France}(2014)}]{Loyd:2014}
{Loyd}, R.~O.~P., \& {France}, K. 2014, \apjs, 211, 9,
  \dodoi{10.1088/0067-0049/211/1/9}

\bibitem[{{Loyd} {et~al.}(2018{\natexlab{a}}){Loyd}, {Shkolnik}, {Schneider},
  {Barman}, {Meadows}, {Pagano}, \& {Peacock}}]{Loyd:2018a}
{Loyd}, R.~O.~P., {Shkolnik}, E.~L., {Schneider}, A.~C., {et~al.}
  2018{\natexlab{a}}, \apj, 867, 70, \dodoi{10.3847/1538-4357/aae2ae}

\bibitem[{{Loyd} {et~al.}(2018{\natexlab{b}}){Loyd}, {France}, {Youngblood},
  {Schneider}, {Brown}, {Hu}, {Segura}, {Linsky}, {Redfield}, {Tian},
  {Rugheimer}, {Miguel}, \& {Froning}}]{Loyd:2018b}
{Loyd}, R.~O.~P., {France}, K., {Youngblood}, A., {et~al.} 2018{\natexlab{b}},
  \apj, 867, 71, \dodoi{10.3847/1538-4357/aae2bd}

\bibitem[{{MacGregor} {et~al.}(2021){MacGregor}, {Weinberger}, {Loyd},
  {Shkolnik}, {Barclay}, {Howard}, {Zic}, {Osten}, {Cranmer}, {Kowalski},
  {Lenc}, {Youngblood}, {Estes}, {Wilner}, {Forbrich}, {Hughes}, {Law},
  {Murphy}, {Boley}, \& {Matthews}}]{MacGregor:2021}
{MacGregor}, M.~A., {Weinberger}, A.~J., {Loyd}, R.~O.~P., {et~al.} 2021,
  \apjl, 911, L25, \dodoi{10.3847/2041-8213/abf14c}

\bibitem[{Marquardt(1963)}]{Marquardt:1963}
Marquardt, D.~W. 1963, Journal of the Society for Industrial and Applied
  Mathematics, 11, 431, \dodoi{10.1137/0111030}

\bibitem[{{Medina} {et~al.}(2022){Medina}, {Charbonneau}, {Winters}, {Irwin},
  \& {Mink}}]{Medina:2022}
{Medina}, A.~A., {Charbonneau}, D., {Winters}, J.~G., {Irwin}, J., \& {Mink},
  J. 2022, arXiv e-prints, arXiv:2203.01344.
\newblock \doarXiv{2203.01344}

\bibitem[{{Newton} {et~al.}(2017){Newton}, {Irwin}, {Charbonneau}, {Berlind},
  {Calkins}, \& {Mink}}]{Newton:2017}
{Newton}, E.~R., {Irwin}, J., {Charbonneau}, D., {et~al.} 2017, \apj, 834, 85,
  \dodoi{10.3847/1538-4357/834/1/85}

\bibitem[{{Newton} {et~al.}(2016){Newton}, {Irwin}, {Charbonneau},
  {Berta-Thompson}, {Dittmann}, \& {West}}]{Newton:2016}
---. 2016, \apj, 821, 93, \dodoi{10.3847/0004-637X/821/2/93}

\bibitem[{{Noyes} {et~al.}(1984){Noyes}, {Hartmann}, {Baliunas}, {Duncan}, \&
  {Vaughan}}]{Noyes:1984}
{Noyes}, R.~W., {Hartmann}, L.~W., {Baliunas}, S.~L., {Duncan}, D.~K., \&
  {Vaughan}, A.~H. 1984, \apj, 279, 763, \dodoi{10.1086/161945}

\bibitem[{{Pineda} {et~al.}(2021{\natexlab{a}}){Pineda}, {Youngblood}, \&
  {France}}]{Pineda:2021a}
{Pineda}, J.~S., {Youngblood}, A., \& {France}, K. 2021{\natexlab{a}}, \apj,
  911, 111, \dodoi{10.3847/1538-4357/abe8d7}

\bibitem[{{Pineda} {et~al.}(2021{\natexlab{b}}){Pineda}, {Youngblood}, \&
  {France}}]{Pineda:2021b}
---. 2021{\natexlab{b}}, \apj, 918, 40, \dodoi{10.3847/1538-4357/ac0aea}

\bibitem[{{Rackham} {et~al.}(2019){Rackham}, {Apai}, \&
  {Giampapa}}]{Rackham:2019}
{Rackham}, B.~V., {Apai}, D., \& {Giampapa}, M.~S. 2019, \aj, 157, 96,
  \dodoi{10.3847/1538-3881/aaf892}

\bibitem[{{Sakaue} \& {Shibata}(2021)}]{Sakaue:2021}
{Sakaue}, T., \& {Shibata}, K. 2021, \apj, 919, 29,
  \dodoi{10.3847/1538-4357/ac0e34}

\bibitem[{{Shields} {et~al.}(2016){Shields}, {Ballard}, \&
  {Johnson}}]{Shields:2016}
{Shields}, A.~L., {Ballard}, S., \& {Johnson}, J.~A. 2016, \physrep, 663, 1,
  \dodoi{10.1016/j.physrep.2016.10.003}

\bibitem[{{Skumanich}(1972)}]{Skumanich:1972}
{Skumanich}, A. 1972, \apj, 171, 565, \dodoi{10.1086/151310}

\bibitem[{{Stauffer} \& {Hartmann}(1986)}]{StaufferHartmann:1986}
{Stauffer}, J.~R., \& {Hartmann}, L.~W. 1986, \apjs, 61, 531,
  \dodoi{10.1086/191123}

\bibitem[{{van der Walt} {et~al.}(2011){van der Walt}, {Colbert}, \&
  {Varoquaux}}]{numpy:2011}
{van der Walt}, S., {Colbert}, S.~C., \& {Varoquaux}, G. 2011, Computing in
  Science and Engineering, 13, 22, \dodoi{10.1109/MCSE.2011.37}

\bibitem[{{Virtanen} {et~al.}(2020){Virtanen}, {Gommers}, {Oliphant},
  {Haberland}, {Reddy}, {Cournapeau}, {Burovski}, {Peterson}, {Weckesser},
  {Bright}, {van der Walt}, {Brett}, {Wilson}, {Millman}, {Mayorov}, {Nelson},
  {Jones}, {Kern}, {Larson}, {Carey}, {Polat}, {Feng}, {Moore}, {VanderPlas},
  {Laxalde}, {Perktold}, {Cimrman}, {Henriksen}, {Quintero}, {Harris},
  {Archibald}, {Ribeiro}, {Pedregosa}, {van Mulbregt}, \& {SciPy 1. 0
  Contributors}}]{scipy:2020}
{Virtanen}, P., {Gommers}, R., {Oliphant}, T.~E., {et~al.} 2020, Nature
  Methods, 17, 261, \dodoi{10.1038/s41592-019-0686-2}

\bibitem[{{Walkowicz} \& {Hawley}(2009)}]{Walkowicz:2009}
{Walkowicz}, L.~M., \& {Hawley}, S.~L. 2009, \aj, 137, 3297,
  \dodoi{10.1088/0004-6256/137/2/3297}

\bibitem[{Waskom(2021)}]{seaborn:2021}
Waskom, M.~L. 2021, Journal of Open Source Software, 6, 3021,
  \dodoi{10.21105/joss.03021}

\bibitem[{{West} {et~al.}(2011){West}, {Morgan}, {Bochanski}, {Andersen},
  {Bell}, {Kowalski}, {Davenport}, {Hawley}, {Schmidt}, {Bernat}, {Hilton},
  {Muirhead}, {Covey}, {Rojas-Ayala}, {Schlawin}, {Gooding}, {Schluns},
  {Dhital}, {Pineda}, \& {Jones}}]{West:2011}
{West}, A.~A., {Morgan}, D.~P., {Bochanski}, J.~J., {et~al.} 2011, \aj, 141,
  97, \dodoi{10.1088/0004-6256/141/3/97}

\bibitem[{{Wilson}(1968)}]{Wilson:1968}
{Wilson}, O.~C. 1968, \apj, 153, 221, \dodoi{10.1086/149652}

\bibitem[{{Woolley}(1936)}]{Woolley:1936}
{Woolley}, R.~V.~D.~R. 1936, \mnras, 96, 515, \dodoi{10.1093/mnras/96.5.515}

\bibitem[{{Wright} {et~al.}(2018){Wright}, {Newton}, {Williams}, {Drake}, \&
  {Yadav}}]{Wright:2018}
{Wright}, N.~J., {Newton}, E.~R., {Williams}, P. K.~G., {Drake}, J.~J., \&
  {Yadav}, R.~K. 2018, \mnras, 479, 2351, \dodoi{10.1093/mnras/sty1670}

\bibitem[{{Youngblood} {et~al.}(2021){Youngblood}, {Pineda}, \&
  {France}}]{Youngblood:2021}
{Youngblood}, A., {Pineda}, J.~S., \& {France}, K. 2021, \apj, 911, 112,
  \dodoi{10.3847/1538-4357/abe8d8}

\end{thebibliography}
\bibliographystyle{aasjournal}



\end{document}